\documentclass[letterpaper,twocolumn,10pt]{article}
\usepackage{usenix}

\usepackage{cite}
\usepackage{graphicx}
\usepackage{xcolor}
\usepackage{listings}
\usepackage{url}
\usepackage{amsmath}
\usepackage{booktabs}
\usepackage{hyperref}  
\usepackage{listings}
\usepackage{xcolor}   
\usepackage{tikz}

\newcommand{\blackcircled}[1]{%
  \tikz[baseline=(char.base)]{
    \node[
      shape=circle,
      fill=black,
      text=white,
      inner sep=0.6pt,
      minimum size=0.8em
    ] (char) {\scriptsize #1};
  }%
}

\usepackage{pifont}

\usepackage{rotating}

\usepackage{xurl} 

\usepackage{xcolor}
\usepackage[colorinlistoftodos,prependcaption,textsize=footnotesize]{todonotes}

\definecolor{michelecolor}{RGB}{255,230,153}   
\definecolor{nicolocolor}{RGB}{189,215,238}    
\definecolor{lorenzocolor}{RGB}{198,239,206}   
\definecolor{rebeccacolor}{RGB}{244,204,204}   

\usepackage{pifont}
\newcommand{\cmark}{\ding{51}}
\newcommand{\xmark}{\ding{55}}

\usepackage{xcolor}
\usepackage{multirow}
\usepackage{float}
\usepackage{verbatim}
\usepackage{placeins}

\makeatletter
\newcommand{\Autoref}[1]{%
  \begingroup
  \def\sectionautorefname{Section}%
\def\subsectionautorefname{Section}%
\def\subsubsectionautorefname{Section}%
  \def\figureautorefname{Figure}%
  \def\tableautorefname{Table}%
  \def\codelistingautorefname{Listing}
  \def\lstlistingautorefname{Listing}
  \autoref{#1}%
  \endgroup
}
\makeatother

\definecolor{backcolour}{rgb}{0.97,0.97,0.97}
\definecolor{codeblue}{rgb}{0.13,0.29,0.53}
\definecolor{codegreen}{rgb}{0.0,0.5,0.0}
\definecolor{codegray}{rgb}{0.5,0.5,0.5}
\definecolor{codeorange}{rgb}{0.8,0.4,0.0}

\lstdefinestyle{cstyle}{
    language=C,
    backgroundcolor=\color{backcolour},
    basicstyle=\small\ttfamily,
    keywordstyle=\bfseries\color{codeblue},
    commentstyle=\itshape\color{codegreen},
    stringstyle=\color{codeorange},
    numberstyle=\tiny\color{codegray},
    numbers=left,
    numbersep=8pt,
    frame=single,
    rulecolor=\color{black!25},
    breaklines=true,
    captionpos=b,
    xleftmargin=\parindent,
    tabsize=4,
}

\newfloat{codelisting}{htbp}{lol}
\floatname{codelisting}{Listing}

\newcommand{\sysname}{\textsf{Sys}}

\usepackage[most]{tcolorbox}
\newcommand{\skel}[1]{%
  \begin{tcolorbox}[
    colback=gray!10,
    boxrule=0pt,
    arc=2pt,
    left=6pt,
    right=6pt,
    top=6pt,
    bottom=6pt
  ]
    \textit{#1}
  \end{tcolorbox}%
}

\newcommand{\tool}{\mbox{\textsf{Antaeus}}}

\newcommand{\toolO}{\mbox{\textsf{Antaeus}\textsuperscript{O4.7}}}

\newcommand{\toolOO}{\mbox{\textsf{Antaeus}\textsuperscript{O4.8}}}

\newcommand{\toolG}{\mbox{\textsf{Antaeus}\textsuperscript{G5.4}}}
\begin{document}

\title{ \Large \bf Antaeus: Hunting Repository-Level Logic Vulnerabilities via Context-Grounded LLM Reasoning}

\author{
{\rm Michele Armillotta}\\
University College London\\
University of Bologna
\and
{\rm Nicolò Romandini}\\
University of Bologna
\and
{\rm Rebecca Montanari}\\
University of Bologna
\and
{\rm Lorenzo Cavallaro}\\
University College London
}
\maketitle

\begin{abstract}
 LLM-based vulnerability detectors have shown promising results in identifying memory-safety bugs and established vulnerability classes, where violations can often be expressed through well-defined security properties such as unsafe data propagation or invalid memory accesses. Logic vulnerabilities, however, pose a different challenge, as their identification requires inferring application-specific security invariants and often relies on implicit assumptions about intended behavior. Even frontier agentic models struggle in this setting, despite their ability to inspect large repositories, because the relevant security invariants are often implicit and buried among large amounts of unrelated code. Motivated by this gap, we present \tool{}, a framework for detecting logic vulnerabilities that grounds LLM reasoning in repository-level code context. \tool{} follows a repository-scale pipeline that combines function prioritization, context-grounded reasoning, comparative validation, and structured reporting. First, it ranks functions using lightweight, repository-wide security signals, directing costly LLM analysis to relevant code regions and reducing model calls and triage effort. For each prioritized function, \tool{} grounds the model in repository evidence, combining local code context with a repository-level view of the application's functionality, security-relevant resources, and trust boundaries. This enables the model to reason about the function within the broader application rather than as an isolated snippet. \tool{} then identifies security-sensitive sinks, derives the safety conditions required for safe execution, and checks whether they are locally satisfied. Candidate findings undergo comparative validation, which prunes concerns reflecting project-wide norms. Finally, \tool{} reports sinks, violated safety conditions, and supporting evidence, making findings specific, actionable, and traceable. We evaluate \tool{} on 35 real-world repositories with confirmed logic vulnerabilities and compare it against function-level LLM analysis and frontier agentic models, including Opus 4.8 Agentic and Codex 5.4. \tool{} detects and explains 20 vulnerabilities, substantially outperforming state-of-the-art baselines while maintaining comparable token usage and cost.

 \end{abstract}

\section{Introduction}
\label{sec:intro}

Logic vulnerabilities arise from failures to enforce application-specific security invariants that govern intended program behavior~\cite{felmetsger2010toward}, and remain an important but underexplored class of defects~\cite{mitre_cwe_top25_2025}. Unlike memory-safety bugs, which manifest through invalid memory accesses, or dataflow vulnerabilities, expressible as source-to-sink propagation, logic vulnerabilities stem from missing, misplaced, or incorrectly composed enforcement of invariants defining who may act, which resources may be accessed, what may be exposed, and under which conditions an action is safe~\cite{mitre_cwe_200, mitre_cwe_284}. Pattern- and taint-based tools such as CodeQL~\cite{codeql-docs} and Semgrep~\cite{semgrep-docs} work when a vulnerability maps to a recurring shape or traceable flow, but logic vulnerabilities typically offer no universal source, sink, or distinguished value to track, so detection instead depends on reconstructing invariants implicitly encoded by the codebase. Project-specific analyses can encode these manually, but doing so requires expertise and does not scale.

Existing work on logic vulnerabilities has made progress, but has largely targeted web applications~\cite{li2013logicscope,sun2014detecting}. White-box approaches~\cite{felmetsger2010toward} infer invariants from normal executions and flag paths that violate them, while black-box approaches~\cite{pellegrino2014toward,deepa2018detlogic} infer behavioral models from HTTP traces and generate tests that break observed patterns. These works confirm that detection depends on recovering intended invariants, but their assumptions stay tied to requests, sessions, forms, and navigation graphs, signals with no analog in repository-level C and C++ software, where invariants are distributed across large codebases and encoded through APIs, callbacks, flags, resource handles, and project conventions: a function may look benign in isolation while its safety depends on conventions, trust boundaries, or checks enforced elsewhere. Detecting logic vulnerabilities here therefore requires recovering invariants from code structure and repository context.

Recent work applies LLMs to vulnerability detection~\cite{guo2025repoaudit,chen2025reasoning}, including agentic systems that inspect repositories and explain findings in natural language, promising because frontier models can reason about program intent and surface real vulnerabilities when properly scaffolded~\cite{AnthropicGlasswing2026, carlini2026mythos}. The limitation is not model capability but how evidence is selected, structured, and budgeted~\cite{li2025everything}: function-level analysis gives too little context, forcing inference from an isolated fragment, while exhaustive exploration supplies that evidence only by spending a large budget on irrelevant code, a strategy that scales poorly and can still bury the facts a candidate operation requires. The issue is context quality and cost, not quantity, and LLM reasoning remains breakable without structured program-semantic support, especially when interprocedural dependencies must be inferred from raw code~\cite{ullah2024llms,yildiz2025benchmarking}. Pipelines that select context, structure reasoning, and validate findings are more effective~\cite{zheng2026veritas}, but success cannot be measured only by whether a bug is eventually found: cost spans candidate generation, validation, impact assessment, and maintainer triage, so cheap candidates can simply shift the bottleneck downstream~\cite{pesoli2026demystifying}. For logic vulnerabilities, this means enough repository evidence to recover application-specific invariants without so much unfiltered code that the model loses focus or exceeds a practical budget.

We argue that repository-level detection requires context grounding: reasoning anchored in explicit evidence about how a project defines principals, resources, trust boundaries, and security-sensitive operations, so as to recover the invariants a vulnerability violates. Without this, model-based detectors risk relying on local syntax or generic priors, insufficient for a property that is inherently project-specific. Grounding is not simply more context: evidence helps only insofar as it explains why a check exists, under what assumptions, and which components jointly enforce the invariant. We therefore frame the problem as invariant recovery rather than retrieval or reasoning: the question is not how much context can be supplied, but whether it lets the model reconstruct the assumptions the repository is meant to operate under, and judge a candidate operation against them.

Based on this observation, we present \tool{}, a repository-level framework for detecting C and C++ logic vulnerabilities through context-grounded model reasoning, which incrementally recovers security invariants from local and project-wide evidence and evaluates each candidate violation against them. Rather than analyzing every function in isolation, \tool{} organizes detection as a staged pipeline: lightweight repository-wide prioritization surfaces functions likely to involve security-relevant behavior; a grounded context is then built for each, combining local code augmentation with a stable repository-level security summary; the model produces structured findings naming security-sensitive sinks, required safety conditions, and whether those conditions are locally satisfied; and comparative validation across similar sinks prunes findings that reflect recurring project-wide norms. The output is a triage-oriented report exposing the sink, the violated condition, and the supporting evidence, so each finding can be inspected and addressed by a reviewer.

We evaluate \tool{} on 35 real-world C and C++ repositories with confirmed CWE-200 and CWE-284 logic vulnerabilities, comparing it against function-level LLM analysis and frontier agentic baselines, including Opus 4.8 Agentic~\cite{anthropic2026opus48} and Codex 5.4~\cite{openai2026gpt54}. \tool{} identifies 20 of 35 vulnerabilities with Opus~4.7 and 14 of 35 with GPT-5.4, in both cases providing structured explanations for the missing conditions and substantially outperforming the baselines. These results show that targeted context grounding recovers repository-specific invariants more effectively than broader agentic exploration, within a comparable token-usage and cost budget.

In summary, we make the following contributions:
\begin{itemize}
    \item We characterize repository-level logic vulnerability detection as evidence-grounded security-invariant checking, where the main challenge is recovering application-specific safety conditions from distributed code context.
    \item We present \tool{}, a staged framework combining repository-wide prioritization, local code augmentation, repository-level security context, structured model reasoning, and comparative validation across analogous code.
    \item We evaluate \tool{} on 35 real-world C and C++ repositories with confirmed CWE-200 and CWE-284 logic vulnerabilities. \tool{} detects 20 of 35 vulnerabilities with Opus~4.7 and 14 of 35 with GPT-5.4, missed by function-level LLM analysis and frontier agentic baselines, while preserving a practical token-usage and cost budget.
\end{itemize}

\section{Motivation}
\label{sec:bg}

\subsection{Why These Classes Resist Detection}
\label{sec:bg:resists}
To illustrate the underlying challenges of logic vulnerability detection, we use CVE-2020-10701, a CWE-284 access-control flaw in libvirt, as a motivating example because it compactly captures the difficulties this class poses. As shown in \Autoref{lst:running}, the pre-patch body of \texttt{virDomainAgentSetResponseTimeout} is the public entry point for configuring how long libvirt waits for QEMU guest-agent replies. It validates the domain handle, retrieves the connection, dispatches to the per-driver callback, and handles errors through the usual path. Every line is locally correct. The vulnerability is the absence of a single guard, \texttt{virCheckReadOnlyGoto(conn->flags, error)}, which would reject read-only callers. Without it, a read-only client can set the timeout to zero and induce a denial-of-service. While a human auditor connects this missing check to libvirt's read-only connection model and the conventions of comparable APIs, automating the same judgment requires recovering invariants that the function body never states. The following subsections detail three challenges that motivate context grounding in \tool{}.
\begin{codelisting}[htbp]
\begin{lstlisting}[style=cstyle,basicstyle=\ttfamily\scriptsize]
int virDomainAgentSetResponseTimeout(virDomainPtr domain, int timeout, unsigned int flags)
{
    virConnectPtr conn;
    VIR_DOMAIN_DEBUG(domain, "timeout=%i, flags=0x%x",
                     timeout, flags);
    virResetLastError();
    virCheckDomainReturn(domain, -1);
    conn = domain->conn;
    if (conn->driver->domainAgentSetResponseTimeout) {
        if (conn->driver->domainAgentSetResponseTimeout(
                domain, timeout, flags) < 0)
            goto error;
        return 0;
    }
    virReportUnsupportedError();
 error:
    virDispatchError(conn);
    return -1;
}
\end{lstlisting}
\caption{Simplified pre-patch body of \texttt{virDomainAgentSetResponseTimeout}
from \texttt{src/libvirt-domain.c} in libvirt (CVE-2020-10701).}
\label{lst:running}
\end{codelisting}
\subsubsection {Non-Local Security Invariants} Many logic vulnerabilities are not confined to the function under analysis. The defining evidence lives in the repository around it. The safety of a function depends not on its local syntax, but on invariants the codebase as a whole is meant to enforce. As \Autoref{lst:running} illustrates, \texttt{virDomainAgentSetResponseTimeout} reads as a standard libvirt API wrapper, and no line within it signals that a read-only check is missing. The violated invariant, that write-privileged public APIs must reject read-only callers, is encoded in libvirt's connection model and in how sibling APIs guard the same boundary, not in any artifact reachable by reading the function alone. Connecting the operation to the policy it violates requires reconstructing repository-level conventions, evidence that local context cannot supply.

\subsubsection{Absence of Syntactic and Dataflow Anchors} Tracking the conditions under which an operation is unsafe requires first knowing which operations are security-sensitive and where to look. For memory-safety and dataflow bugs, a detector has a stable scaffolding. There is a dangerous operation to flag, a tainted source to trace from, a sink to trace to, and a flow that links them. For logic bugs, none of these anchors exists. In \Autoref{lst:running}, there is no tainted input to track and no dataflow trail that walks toward the defect. The security-relevant operation is a configuration mutation that changes an agent timeout, which becomes sensitive only once the repository's classification of it as a write operation is recovered. The evidence required to judge the function is thus a semantic classification the codebase assigns rather than a syntactic property a detector can pattern-match.

\subsubsection{Distinguishing Violations from Project Norms} 
\label{sec:why_compare}
Detecting a logic vulnerability requires determining not only that a safety condition is unmet, but whether the unmet condition is a genuine anomaly. An LLM model evaluating a function locally is usually assertive~\cite{10.1109/ICSE55347.2025.00038, turpin2023language}. Shown a sensitive-looking operation, it tends to raise the safety conditions appropriate to that kind of operation, regardless of whether the surrounding codebase already addresses them. Across a repository, this yields a characteristic false-positive pattern, in which the same unsatisfied condition is reported for many structurally similar sinks, often in similar phrasing. A genuine logic vulnerability, by contrast, is an anomaly~\cite{10.1145/502034.502041,10.1145/2508859.2516665}, a deviation from how comparable code in the same project handles the same operation. When many similar operations are handled uniformly, that uniformity is the norm, and a model that raises the same concern across all of them is flagging the norm, not a distinctive violation. As \Autoref{lst:running} shows, the missing read-only check is meaningful only relative to the sibling APIs that do enforce the boundary. A suspected violation, therefore, remains only a hypothesis until it can be set against the project's enforced pattern.

\subsection{Insights for Context-Grounded Logic Vulnerability Detection}
\label{sec:insights}
The challenges above motivate context grounding, which constrains model reasoning to evidence on how a repository defines its security invariants rather than letting the model fall back on local syntax and generic priors. Grounding combines two complementary layers. Local program evidence ties claims about a function to concrete repository artifacts, and repository-level context supplies the application-specific invariants against which a local operation must be judged. This leads to three design insights.
\underline{Insight \blackcircled{1}:} Prioritize candidates before grounding. Because logic vulnerabilities admit no syntactic anchor, every function is a nominal candidate, but grounding each one is impractical. Detection should begin with repository-wide prioritization that surfaces functions likely to involve security-relevant behavior, spending the analysis budget only where a policy violation could plausibly occur. \underline{Insight \blackcircled{2}:} Pair local evidence with repository context. The information needed to judge a function lies outside it, so the model should reason over both layers simultaneously. Local code, callees, macros, constants, and types anchor claims to concrete artifacts, while a repository-level security summary of the application's purpose, protected objects, and trust topology supplies the invariants the operation must satisfy.
\underline{Insight \blackcircled{3}:} A missing safety condition is meaningful only relative to how similar code in the same repository handles the same operation. Each flagged condition should be validated against analogous sinks, discarding those that recur uniformly as project norms and retaining those that set the target apart.
These insights shape the design of \tool{}, a framework that grounds LLM-based logic vulnerability detection in repository evidence. A repository-wide prioritizer narrows the candidate set to security-relevant functions; a grounding stage pairs local code augmentation with a repository-level security summary; a structured reasoning stage produces findings in terms of security-sensitive sinks and their required safety conditions; and a comparative validator confirms genuine violations by setting each finding against analogous code.

\subsection{Threat model}
\label{sec:bg:Threat}
We consider an adversary who interacts with the target software through its intended interfaces, such as API calls, configuration options, files, command-line arguments, or network requests. The adversary cannot modify the program code, but can choose inputs, parameters, call sequences, and execution contexts to violate security-relevant assumptions made by the application. We focus on logic vulnerabilities in which the executed code is syntactically valid, but an operation is performed under semantically unsafe conditions.
\tool{} assumes access to the source repository under analysis, but does not assume exploit inputs, runtime traces, or an execution environment. Its goal is to statically detect and explain candidate logic vulnerabilities by linking security-sensitive code to the repository context that determines whether its use is safe. Memory-safety bugs and syntactic defects are out of scope.

\begin{figure*}[t]
  \centering
  \includegraphics[width=\textwidth]{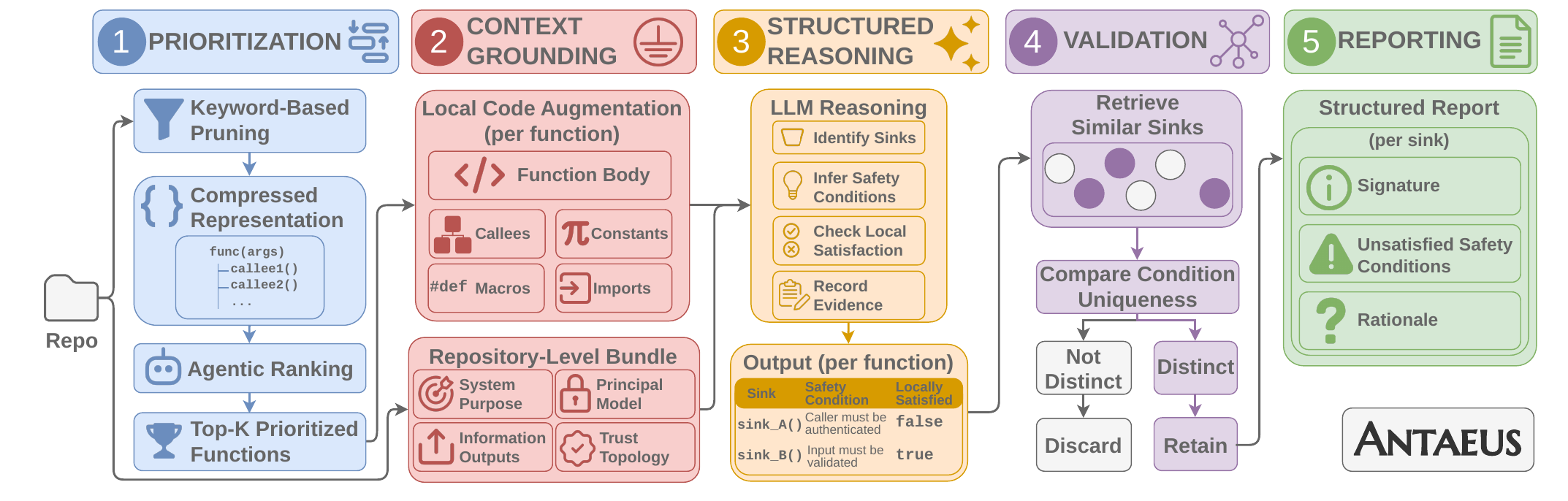}
  \caption{Overview of the \tool{} pipeline:
  prioritization, context-grounding, structured reasoning,
  validation, and reporting.}
  \label{fig:pipeline}
\end{figure*}
\section{Design}
\label{sec:design}

The design follows directly from the insights identified in \Autoref{sec:insights}. \tool{} is a five-stage repository-level pipeline, shown in \Autoref{fig:pipeline}: function prioritization, context-grounding, structured reasoning, comparative validation, and reporting. These stages also reflect recent state-of-the-art practice in LLM-based vulnerability detection~\cite{carlini2026mythos, zheng2026veritas}. First, function prioritization ranks functions using lightweight repository-wide security signals, focusing LLM analysis on the most relevant code regions. Second, context-grounding supplies the model with local code augmentation and a repository-level security summary, tying its reasoning to concrete repository evidence. Third, structured reasoning prompts the model to identify security-sensitive sinks together with the safety conditions required for their safe use, and to record whether each condition is locally satisfied. Fourth, comparative validation checks whether a flagged condition is distinctive with respect to similar sinks in the same repository, pruning conditions that recur as project norms. Finally, reporting presents the retained findings as actionable, evidence-grounded explanations. The remainder of this section examines each stage in detail.

\subsection{Prioritization}
\label{sec:design:prioritization}

At repository scale, the first challenge is deciding where to apply expensive LLM analysis. As discussed in \Autoref{sec:insights}, exhaustive repository exploration spends most of the budget on irrelevant code, while letting an agentic model search unaided does not guarantee that every security-relevant function is examined and may lead it to rely on superficial cues.

\tool{} therefore begins with a lightweight, high-recall prioritization stage. The goal of this stage is not to classify functions as vulnerable, but to reduce the repository to a ranked set of functions that are most likely to contain logic-relevant behavior. This lets the later LLM stages operate at the repository level while keeping the number of detailed analyses tractable. The prioritization stage first prunes the codebase, retaining only C/C++ source and header files. It then applies broad keyword-based heuristics to remove functions that are unlikely to involve authorization, access control, trust-boundary enforcement, configuration changes, or other security-sensitive operations. These heuristics are intentionally loose: they are designed to remove clearly irrelevant functions while preserving recall for functions that expose even weak security signals. After this initial pruning, \tool{} compresses the remaining repository into a lightweight function-level representation. For each retained function, it removes the internal implementation logic and keeps only the function signature together with the callees invoked inside the function body. This abstraction is deliberately simple. It discards local implementation detail, but preserves the repository-wide call and naming signals that are useful for identifying security-relevant code regions. In particular, it allows the model to see many more functions and callees than would fit if full function bodies were provided. The compressed repository is then split into chunks that fit within the model's context budget, with each chunk represented as a single line aggregating the functions it contains. These chunks are passed to an agentic LLM, which ranks the functions by their relevance to the target logic vulnerability classes.

This design serves two purposes. First, it gives the prioritization model a global view of the repository's security signals. Rather than inspecting a small subset of files or relying on its training-time knowledge of popular projects, the model receives a compact representation of the functions that remain after high-recall pruning. This makes it more likely to identify functions whose names, callees, or surrounding API structure indicate relevance to CWE-284 or CWE-200. Second, it limits the cost of the subsequent analysis. The detailed context-grounded reasoning stage is applied only to prioritized functions, reducing model calls, token usage, and the number of candidate reports that must be reviewed by a human auditor.

\subsection{Contextual Grounding}
\label{sec:design:grounding}

The contextual grounding stage constructs the evidence needed to interpret a function in its repository context. The unit of analysis is a single function, but a function does not stand alone: its meaning depends on the types of values it receives, the bodies of the functions it calls, the expansion of the macros it uses, the constants it references, and the headers and imports that determine the names in scope. Asking an LLM to reason about a function in isolation forces it to fill in these blanks from general knowledge or from patterns seen during training, which is unreliable in real C/C++ projects: a project-specific macro may not expand to what the model assumes, and a constant may be defined to a value that changes the meaning of a check. For this reason, each function is enriched with a local augmentation bundle extracted from the repository via a lightweight Tree-sitter static analysis pass. The parser identifies function definitions, call sites, macro and constant references, and include relationships, from which we build a repository-level graph linking each analyzed function to the artifacts needed to interpret it, an efficient index for retrieving nearby evidence rather than a complete semantic model of the program.

The local augmentation bundle contains the signatures of direct callees at depth one, the bodies of those callees when defined within the project, the textual definitions of referenced macros and constants, and the relevant typedefs and imports in scope, supplied as explicit program evidence against which claims about identifiers, callees, or constants can be cross-checked. The choice of depth one is deliberate: expanding the call graph further increases prompt size and dilutes the model's attention without guaranteeing complete interprocedural coverage, while one level is sufficient to disambiguate the function's immediate semantics and its invoked project APIs within budget. Deeper semantic dependencies are captured, when relevant, through the repository context bundle.

Local augmentation pinpoints what each line of the function does, but not what the program is for, and CWE-284 and CWE-200 vulnerabilities depend on exactly that repository-level semantics, none of which is reliably recoverable from a single function even with depth-one expansion. We therefore compute, once per repository, a structured context bundle produced by a repository-level LLM agent before function-level analysis. Given the complete repository, the agent extracts a stable description of the project's purpose, actors, resources, outputs, and exposed interfaces, not to decide whether any function is vulnerable but to give subsequent function-level prompts a shared repository-level frame. Each item is associated with repository-relative paths and short code excerpts, so later reasoning can be tied back to concrete evidence.

The bundle contains five sections. The system purpose section describes what kind of artifact the repository implements, its deployment context, and its primary consumers. The principal model section identifies the actors or callers the code distinguishes, how their identity is established, and through which interfaces they reach the system. The protected objects section records security-relevant repository objects, such as configuration state, service handles, credentials, filesystem paths, or shared state tables, without deciding which principals may access them. The information outputs section summarizes externally observable outputs, including functional responses, errors, status information, configuration data, and internal state exposed through return paths or response builders. Finally, the trust topology section records the interfaces exposed by the project and the structural boundaries they create, such as library APIs, IPC channels, local sockets, filesystem interfaces, or network-facing entry points.

\subsection{Structured LLM Reasoning}
\label{sec:design:reasoning}

The structured LLM reasoning stage uses the grounded input to produce a function-level finding, not an unconstrained vulnerability label: the model is queried for a fixed-format finding whose elements must be tied to observable code or explicit repository evidence, aligning with prior observations that explanations should rest on verifiable evidence rather than unconstrained post-hoc rationalizations~\cite{nadeem2023sok}. This format in turn determines what contextual grounding must supply: enough local and repository-level evidence for the model to identify security-sensitive operations and state whether the required safety properties hold.

For each function, the model first identifies a set of \emph{sinks}, operations syntactically present in the function whose execution is potentially sensitive to the CWE under consideration. For CWE-284, a sink is any operation affecting a protected resource, such as a privileged system call, a write to a resource exposed across a trust boundary, or a state update performed on behalf of a principal. For CWE-200, a sink is any operation that emits data across a trust boundary into a region observable by a less privileged principal, such as returning data through an API, writing to a network channel, logging sensitive information, or exposing internal state through an externally visible interface.

For each sink, the model then derives a set of \emph{safety conditions}: predicates over the surrounding code, local checks, or calling context that must hold for the sink to be safe in this repository. Each carries a \texttt{locally\_satisfied} flag and a short justification, both tied to specific code in the function or to material supplied by contextual grounding. A function is reported as a candidate vulnerability when at least one safety condition of at least one sink is not locally satisfied.

This scheme serves three purposes. First, it anchors reasoning in concrete evidence: by committing to specific sinks and checkable predicates, the model cannot hide a vague intuition behind generic prose, mitigating the unfaithfulness of free-form LLM rationales. Second, the output is directly useful to an auditor, who is told which operation is security-sensitive and which predicate to verify, rather than handed a binary verdict to validate from scratch. Third, the format enables sharper evaluation: a finding is accepted only when both the judgment and its supporting reasoning are accurate, so findings with a correct label but evidence inconsistent with the grounded material are discarded as incorrectly reasoned.

\subsection{Comparative Validation}
\label{sec:design:validation}
The comparative validation stage operationalizes the observation in \Autoref{sec:why_compare}, that a genuine logic vulnerability is an anomaly relative to its peers while a recurring concern reflects a project norm. It reframes false-positive elimination as anomaly detection over the structured findings produced by the LLM. For each flagged condition, the stage compares the target against structurally similar sinks in the same repository, retaining the condition when it sets the target apart from its peers and discarding it when the same concern recurs uniformly across them.

To decide whether an unsatisfied safety condition is distinctive or merely a
recurring pattern, the validation stage compares the target sink against similar
sinks from the same repository. Similarity is computed in two steps. First, each
sink is represented by its \emph{sink identifier}, i.e., the code fragment that
corresponds to the security-sensitive operation reported by the model. We embed
these sink identifiers with UniXcoder \cite{guo2022unixcoder} and use cosine similarity to build a
neighborhood of sinks that are syntactically or semantically close at the code
level. This step compares operations, not model explanations. Second, validation compares the unsatisfied safety conditions attached to those
sinks. Safety conditions are natural-language predicates produced by the model,
so we embed them with all-MiniLM-L6-v2 sentence-transformer model and compare them using cosine
similarity. A target condition is considered repeated when enough neighboring
sinks have an unsatisfied condition that is semantically similar to it. In that
case, the condition is treated as a repository-wide recurring pattern.

Both comparisons require a notion of \emph{similar enough}. We avoid fixed global
thresholds because similarity distributions vary across repositories: some
codebases contain many near-duplicate API wrappers, while others are more
heterogeneous. We therefore calibrate thresholds per repository using the
empirical distribution of similarities observed in that repository. For a sample
$S$, each threshold is defined as $\tau = \mu(S) + n \cdot \sigma(S)$.
The strictness parameter $n$ is set independently for each threshold. Lower values favor recall by accepting more neighbors and more repeated conditions; higher values make pruning more conservative.

The validation stage instantiates four thresholds:

\begin{itemize}
    \item $\tau_{\mathit{sink}}$ is the minimum UniXcoder cosine similarity for two sink identifiers to be considered neighboring sinks. Its sample $S$ is the set of similarities over sink pairs in the repository.

    \item $\tau_{\mathit{cond}}$ is the minimum sentence-transformer cosine
    similarity for two unsatisfied safety conditions to be considered the same
    condition. Its sample $S$ is the set of similarities over condition pairs in
    the repository.

    \item $\tau_{\mathit{maj}}$ is the minimum fraction of neighboring sinks
    that must contain a similar unsatisfied condition for the target condition
    to be considered recurring.

    \item $\tau_{\mathit{min}}$ is the minimum number of neighboring sinks
    required before a pruning decision is allowed.
\end{itemize}

These thresholds define a single pruning rule. For a target sink $s$ with an
unsatisfied safety condition $c$, let $N(s)$ be the set of neighboring sinks
selected using $\tau_{\mathit{sink}}$. Let $C(v)$ be the set of unsatisfied
safety conditions reported for a neighboring sink $v$. The coverage of $c$ is
the fraction of neighboring sinks that contain at least one sufficiently similar
condition:
{\small
\begin{equation}
\mathrm{cov}(c) =
\frac{
|\{v \in N(s) \mid \exists c_v \in C(v):
\mathrm{sim}_{\mathit{cond}}(c,c_v) \geq \tau_{\mathit{cond}}\}|
}{
|N(s)|
}.
\end{equation}
}
The condition is pruned when the neighborhood is large enough, and the condition is repeated across a sufficient fraction of that neighborhood, i.e, $
\mathrm{prune}(c) \equiv
|N(s)| \geq \tau_{\mathit{min}}
\wedge
\mathrm{cov}(c) \geq \tau_{\mathit{maj}}.
$

Intuitively, a finding is retained when its unsatisfied condition is anomalous
among similar sinks. If many similar sink identifiers in the same repository
trigger the same unsatisfied condition, the condition is likely to reflect a
model bias or a project-wide coding pattern.
If the condition appears only at the target sink, or only in too small a
neighborhood, validation keeps it for reporting. This repository-local
comparison gives \tool{} a signal that isolated function analysis cannot use:
whether the alleged violation is distinctive relative to analogous operations in
the same codebase.

\subsection{Reporting}
\label{sec:design:reporting}

The final stage emits the findings retained after comparative validation as structured, evidence-grounded reports. The report contains one entry per flagged function, where a function is flagged when it contains at least one security-sensitive sink with a required condition that is not locally satisfied. Functions whose sinks all meet their required conditions are not reported. This structure makes every claim auditable against concrete repository artifacts and exposes the results in a machine-readable form suitable for downstream triage and aggregation.
Each reported entry identifies the analyzed function by its name, file, line range, and a function \texttt{id}, and records the security-sensitive sinks identified within it. For each sink, \tool{} reports the operation, a short description of the relevant trust or exposure boundary, and the conditions required for the operation to be secure. Each condition specifies the expected invariant, whether it is locally satisfied, and a code-grounded justification for that judgment. An unsatisfied condition, together with its justification, constitutes the vulnerability evidence. Accordingly, a function is included in the final report only if at least one required condition is not locally satisfied. The complete structure of a report entry is provided in Appendix~\ref{app:report-schema}.

This design is consistent with recommendations from the explainable security literature, which stresses that explanations for security applications should support analyst decision-making while remaining auditable, understandable, and grounded in the evidence used to produce them~\cite{nadeem2023sok}.

\section{Evaluation}
\label{sec:eval}

The central question is whether structuring repository-level context the way \tool{} does actually helps an LLM uncover logic vulnerabilities, and whether the gains are worth their cost. We break this down into four research questions:

\begin{itemize}
    \item \textbf{RQ1 (Effectiveness).} Can a multi-step LLM framework detect logic vulnerabilities at the repository level?
    \item \textbf{RQ2 (Comparison).} How does \tool{} compare against state-of-the-art agentic LLMs and function-level baselines?
    \item \textbf{RQ3 (Ablation).} Which components
    drive the gains, and by how much?
    \item \textbf{RQ4 (Cost).} What is the cost of running \tool{}, and how does it compare to alternatives?
\end{itemize}

\subsection{Experimental Setup}

\noindent\textbf{Datasets:} {\color{black} We draw our evaluation targets primarily from ReposVul~\cite{wang2024reposvul}, a repository-level vulnerability dataset built from NVD CVE records and their patch commits, spanning thousands of open-source projects across multiple languages. ReposVul annotates each entry at the repository, file, function, and line levels, and ships every CVE with its full repository, allowing us to analyze each vulnerability in its real project context. From ReposVul, we select 28 CVEs, comprising 12 CWE-200 and 16 CWE-284 logic vulnerabilities, each paired with its repository. We further construct a post-cutoff evaluation set by identifying 7 additional repository-level CVEs through targeted Web scraping and manual Internet search. These vulnerabilities, comprising 3 CWE-200 and 4 CWE-284 cases, were disclosed after March 2026, which is later than the knowledge cutoff of all models used in our experiments. Overall, our evaluation comprises 35 repository-level vulnerability cases. The selection was randomized, with the sample size limited by budget and by the need for reliable ground truth.}

\noindent \textbf{Models:} We run \tool{} with two frontier models and refer to the resulting configurations by the reasoning model they use: \toolO{} (Claude Opus~4.7) and \toolG{} (GPT-5.4). We additionally report recall for \toolOO{} (Claude Opus~4.8). All analyses run through model APIs and require no local compute.

\noindent\textbf{Evaluation Procedure:} A repository is given to \tool{} without any prior indication of which vulnerability, if any, it contains, since in a realistic setting, the class of a latent flaw is unknown. To reflect this, we run \tool{} over each repository in two independent passes, one targeting CWE-284 and one targeting CWE-200. Each pass instantiates the pipeline with the sink and safety-condition definitions appropriate to its class, so that prioritization, structured reasoning, and validation are all specialized to the vulnerability type under consideration. The two passes are otherwise identical and operate on the same repository.
\textcolor{black}{%
Because LLM-based pipelines are inherently non-deterministic, we repeat the full evaluation three times for \tool{} and for every baseline, to obtain a measure of the stability and reproducibility of our results.
}

\noindent \textbf{Metrics:} \textcolor{black}{%
We report the number of true positives (TPs), annotated with a stability count indicating in how many of the three runs each TP was recovered, and the number of false positives (FPs), averaged across the three runs.
}Scoring recall for logic vulnerabilities is nontrivial, since a tool may flag the right function for the wrong reason, and a binary vulnerable versus non-vulnerable match would overstate how well the underlying flaw was understood. We therefore exploit \tool{}'s structured output. Because each finding names a security-sensitive sink together with the conditions under which it is safe, we count a detection as a true positive only when the reported sink and its conditions actually explain the CVE. Concretely, a vulnerability is counted as detected when the pass corresponding to its CWE class reports a finding that explains it, while the opposing pass contributes only to the false-positive count for that repository. The number of false positives is then the count of reported findings that do not correspond to a real flaw. \textcolor{black}{%
 To make this judgment explicit, we apply a two-part labeling protocol. First, we require a location match, where the reported function and vulnerability-relevant operation correspond to the location and operation implicated by the CVE. Second, we require a concept match, meaning that the security condition and underlying vulnerability described by the finding are consistent with the flaw documented in the CVE. We determine concept match through independent inspection and consensus among the authors, since CVE descriptions and model explanations may use substantially different terminology while describing the same underlying flaw. Finally, because each ReposVul instance establishes the presence of a known CVE but does not establish the absence of additional CWE-200/CWE-284 vulnerabilities in the repository, our reported counts should be interpreted conservatively. The TP count is a lower bound, since we do not attempt to identify additional vulnerabilities beyond the ground-truth CVE, while the FP count is an upper bound, since findings labeled as false positives may in principle correspond to previously undocumented vulnerabilities. Thus, our evaluation represents a worst-case assessment with respect to both missed true positives and potentially overestimated false positives.
}

\noindent \textbf{Comparisons:} We compare \tool{} against two families of baselines. The first is \emph{function-level detection}, where each function is analyzed in isolation with no repository-level context, using the same models as \tool{} (Opus 4.7 or GPT-5.4) to separate the contribution of context from that of the underlying model. The second is \emph{agentic detection}, where a model autonomously explores the repository, for which we use Claude Opus 4.8, Claude Opus 4.7, and Codex 5.4.

\subsection{Prioritization Statistics}
\label{sec:prioritization}

\begin{table}[t]
\centering
\caption{Aggregate pruning statistics across the 35 repositories. \emph{Ret.} denotes files retained after heuristic pruning; \emph{Scope} and \emph{Anal.} denote in-scope and analyzed functions; \emph{Red.} is their reduction ratio.}
\label{tab:pruning-summary}
\renewcommand{\arraystretch}{0.9}

\begin{tabular}{l r r r r r}
\toprule
Pass & Files & Ret. & Scope & Anal. & Red. \\
\midrule
284 & 54{,}462 & 6{,}668  & 85{,}332  & 2{,}977 & 28.7$\times$ \\
200 & 54{,}462 & 12{,}547 & 101{,}527 & 3{,}135 & 32.4$\times$ \\
\midrule
\textbf{Tot.}
& \textbf{--}
& \textbf{19{,}215}
& \textbf{186{,}859}
& \textbf{6{,}112}
& \textbf{30.6$\times$} \\
\bottomrule
\end{tabular}

\end{table}

Both passes operate on a prioritized subset of each codebase. \Autoref{tab:pruning-summary} reports the aggregate effect of this stage across the 35 repositories. Starting from 54{,}462 source and header files, heuristic pruning and agent ranking reduce the in-scope functions to the set actually analyzed, yielding roughly a 29$\times$ reduction for the CWE-284 pass and a 33$\times$ reduction for the CWE-200 pass. \textcolor{black}{%
Despite this aggressive reduction, every ground-truth vulnerable function is retained in the prioritized set, corresponding to a prioritization recall of 100\%.
} Detailed LLM reasoning is therefore applied to a few thousand prioritized functions, while still recovering the majority of detected vulnerabilities. That detection holds under such a reduced budget indicates that prioritization preserves the relevant security signal while bounding the amount of code that must be analyzed and later triaged. The per-repository breakdown is given in Appendix~\ref{app:pruning}.

\begin{table*}[!t]
\centering 

\caption{Comparison of \toolO{} with Claude-based function-level and agentic (Ag.) baselines, with and without ranking (Rk.). Subscripts indicate stability across three runs; FPs are means with observed ranges. IDs refer to Table ~\ref{tab:id-mapping}.}
\label{tab:comparison-claude}
\setlength{\tabcolsep}{6pt}                 
\renewcommand{\arraystretch}{0.9}
\begin{tabular}{llrlrlrlrlrlr}

\toprule
& \multicolumn{2}{c}{\toolO{}} & \multicolumn{2}{c}{Func.} & \multicolumn{2}{c}{4.7 Ag.} & \multicolumn{2}{c}{4.7 Ag.+Rk.} & \multicolumn{2}{c}{4.8 Ag.} & \multicolumn{2}{c}{4.8 Ag.+Rk.} \\
\cmidrule(lr){2-3}\cmidrule(lr){4-5}\cmidrule(lr){6-7}\cmidrule(lr){8-9}\cmidrule(lr){10-11}\cmidrule(lr){12-13}
ID & TPs & FPs & TPs & FPs & TPs & FPs & TPs & FPs & TPs & FPs & TPs & FPs \\
\midrule
01 & \cmark$_{2/3}$ & $80_{[71,86]}$ & \xmark$_{0/3}$ & $188_{[130,218]}$ & \xmark$_{0/3}$ & $12_{[0,23]}$ & \xmark$_{0/3}$ & $80_{[58,104]}$ & \xmark$_{0/3}$ & $15_{[13,16]}$ & \xmark$_{0/3}$ & $31_{[22,40]}$ \\
02 & \cmark$_{2/3}$ & $35_{[33,37]}$ & \xmark$_{0/3}$ & $39_{[37,44]}$ & \xmark$_{0/3}$ & $16_{[0,30]}$ & \cmark$_{1/3}$ & $26_{[25,26]}$ & \xmark$_{0/3}$ & $15_{[12,18]}$ & \xmark$_{0/3}$ & $24_{[23,26]}$ \\
03 & \xmark$_{0/3}$ & $49_{[43,53]}$ & \xmark$_{0/3}$ & $53_{[46,58]}$ & \xmark$_{0/3}$ & $10_{[0,15]}$ & \xmark$_{0/3}$ & $31_{[19,43]}$ & \xmark$_{0/3}$ & $10_{[4,14]}$ & \xmark$_{0/3}$ & $24_{[17,31]}$ \\
04 & \xmark$_{0/3}$ & $6_{[4,7]}$ & \xmark$_{0/3}$ & $15_{[6,20]}$ & \xmark$_{0/3}$ & $9_{[0,19]}$ & \xmark$_{0/3}$ & $31_{[13,42]}$ & \xmark$_{0/3}$ & $3_{[1,6]}$ & \xmark$_{0/3}$ & $17_{[11,27]}$ \\
05 & \cmark$_{3/3}$ & $342_{[319,356]}$ & \cmark$_{3/3}$ & $314_{[231,362]}$ & \xmark$_{0/3}$ & $12_{[0,28]}$ & \cmark$_{1/3}$ & $88_{[53,140]}$ & \cmark$_{1/3}$ & $9_{[7,12]}$ & \cmark$_{1/3}$ & $60_{[18,103]}$ \\
06 & \xmark$_{0/3}$ & $22_{[21,23]}$ & \xmark$_{0/3}$ & $9_{[6,11]}$ & \xmark$_{0/3}$ & $12_{[0,27]}$ & \xmark$_{0/3}$ & $26_{[17,36]}$ & \xmark$_{0/3}$ & $7_{[5,8]}$ & \xmark$_{0/3}$ & $16_{[8,21]}$ \\
07 & \cmark$_{1/3}$ & $1_{[1,1]}$ & \xmark$_{0/3}$ & $2_{[0,4]}$ & \xmark$_{0/3}$ & $2_{[0,6]}$ & \xmark$_{0/3}$ & $5_{[2,10]}$ & \cmark$_{1/3}$ & $1_{[0,3]}$ & \xmark$_{0/3}$ & $5_{[2,7]}$ \\
08 & \cmark$_{3/3}$ & $6_{[5,7]}$ & \cmark$_{1/3}$ & $9_{[9,9]}$ & \cmark$_{1/3}$ & $5_{[0,12]}$ & \cmark$_{1/3}$ & $13_{[10,16]}$ & \xmark$_{0/3}$ & $4_{[4,5]}$ & \cmark$_{1/3}$ & $7_{[7,7]}$ \\
09 & \cmark$_{3/3}$ & $23_{[17,28]}$ & \cmark$_{3/3}$ & $16_{[9,21]}$ & \cmark$_{2/3}$ & $8_{[0,13]}$ & \cmark$_{2/3}$ & $32_{[23,44]}$ & \cmark$_{3/3}$ & $5_{[5,6]}$ & \cmark$_{1/3}$ & $13_{[11,16]}$ \\
10 & \cmark$_{3/3}$ & $103_{[94,108]}$ & \xmark$_{0/3}$ & $54_{[38,74]}$ & \xmark$_{0/3}$ & $9_{[0,17]}$ & \cmark$_{1/3}$ & $40_{[36,43]}$ & \xmark$_{0/3}$ & $3_{[2,4]}$ & \xmark$_{0/3}$ & $7_{[5,9]}$ \\
11 & \cmark$_{3/3}$ & $79_{[77,83]}$ & \xmark$_{0/3}$ & $50_{[35,60]}$ & \xmark$_{0/3}$ & $8_{[0,18]}$ & \cmark$_{2/3}$ & $44_{[13,69]}$ & \cmark$_{2/3}$ & $6_{[4,8]}$ & \xmark$_{0/3}$ & $16_{[12,19]}$ \\
12 & \xmark$_{0/3}$ & $25_{[24,28]}$ & \xmark$_{0/3}$ & $47_{[25,63]}$ & \xmark$_{0/3}$ & $10_{[0,19]}$ & \xmark$_{0/3}$ & $10_{[3,16]}$ & \xmark$_{0/3}$ & $6_{[3,11]}$ & \xmark$_{0/3}$ & $10_{[8,11]}$ \\
13 & \xmark$_{0/3}$ & $49_{[46,51]}$ & \xmark$_{0/3}$ & $99_{[48,130]}$ & \xmark$_{0/3}$ & $8_{[0,18]}$ & \xmark$_{0/3}$ & $49_{[18,96]}$ & \xmark$_{0/3}$ & $9_{[9,10]}$ & \xmark$_{0/3}$ & $16_{[14,17]}$ \\
14 & \xmark$_{0/3}$ & $29_{[28,30]}$ & \xmark$_{0/3}$ & $91_{[38,122]}$ & \xmark$_{0/3}$ & $10_{[0,23]}$ & \xmark$_{0/3}$ & $30_{[13,64]}$ & \xmark$_{0/3}$ & $7_{[5,8]}$ & \xmark$_{0/3}$ & $7_{[6,8]}$ \\
15 & \cmark$_{3/3}$ & $3_{[2,4]}$ & \cmark$_{2/3}$ & $19_{[6,26]}$ & \cmark$_{1/3}$ & $8_{[0,15]}$ & \cmark$_{2/3}$ & $12_{[7,15]}$ & \cmark$_{3/3}$ & $2_{[1,5]}$ & \cmark$_{2/3}$ & $9_{[7,10]}$ \\
16 & \cmark$_{1/3}$ & $5_{[5,6]}$ & \cmark$_{3/3}$ & $18_{[8,23]}$ & \xmark$_{0/3}$ & $5_{[0,14]}$ & \xmark$_{0/3}$ & $7_{[6,10]}$ & \xmark$_{0/3}$ & $4_{[2,7]}$ & \xmark$_{0/3}$ & $6_{[3,8]}$ \\
17 & \xmark$_{0/3}$ & $41_{[37,44]}$ & \xmark$_{0/3}$ & $52_{[35,61]}$ & \xmark$_{0/3}$ & $6_{[0,15]}$ & \xmark$_{0/3}$ & $16_{[10,24]}$ & \xmark$_{0/3}$ & $4_{[2,5]}$ & \xmark$_{0/3}$ & $7_{[4,11]}$ \\
18 & \xmark$_{0/3}$ & $9_{[7,13]}$ & \xmark$_{0/3}$ & $0_{[0,0]}$ & \xmark$_{0/3}$ & $6_{[0,13]}$ & \xmark$_{0/3}$ & $8_{[4,14]}$ & \xmark$_{0/3}$ & $3_{[2,4]}$ & \xmark$_{0/3}$ & $5_{[5,6]}$ \\
19 & \cmark$_{3/3}$ & $54_{[47,58]}$ & \cmark$_{2/3}$ & $85_{[64,98]}$ & \xmark$_{0/3}$ & $12_{[0,26]}$ & \xmark$_{0/3}$ & $35_{[24,54]}$ & \xmark$_{0/3}$ & $8_{[7,8]}$ & \xmark$_{0/3}$ & $16_{[14,18]}$ \\
20 & \cmark$_{3/3}$ & $20_{[15,24]}$ & \xmark$_{0/3}$ & $4_{[4,5]}$ & \xmark$_{0/3}$ & $11_{[0,21]}$ & \cmark$_{1/3}$ & $10_{[6,14]}$ & \cmark$_{1/3}$ & $4_{[1,7]}$ & \cmark$_{1/3}$ & $5_{[5,6]}$ \\
21 & \cmark$_{3/3}$ & $79_{[74,82]}$ & \xmark$_{0/3}$ & $200_{[64,275]}$ & \xmark$_{0/3}$ & $11_{[0,17]}$ & \xmark$_{0/3}$ & $37_{[34,44]}$ & \xmark$_{0/3}$ & $7_{[3,9]}$ & \xmark$_{0/3}$ & $14_{[12,18]}$ \\
22 & \xmark$_{0/3}$ & $52_{[46,57]}$ & \xmark$_{0/3}$ & $46_{[28,57]}$ & \xmark$_{0/3}$ & $13_{[0,22]}$ & \xmark$_{0/3}$ & $52_{[41,72]}$ & \xmark$_{0/3}$ & $15_{[13,18]}$ & \xmark$_{0/3}$ & $32_{[20,44]}$ \\
23 & \xmark$_{0/3}$ & $25_{[21,29]}$ & \cmark$_{2/3}$ & $35_{[25,40]}$ & \xmark$_{0/3}$ & $10_{[0,20]}$ & \xmark$_{0/3}$ & $17_{[16,18]}$ & \xmark$_{0/3}$ & $5_{[4,7]}$ & \xmark$_{0/3}$ & $9_{[6,14]}$ \\
24 & \cmark$_{3/3}$ & $92_{[90,95]}$ & \cmark$_{3/3}$ & $186_{[173,195]}$ & \xmark$_{0/3}$ & $17_{[0,30]}$ & \xmark$_{0/3}$ & $48_{[38,58]}$ & \xmark$_{0/3}$ & $4_{[3,5]}$ & \xmark$_{0/3}$ & $12_{[10,16]}$ \\
25 & \xmark$_{0/3}$ & $94_{[82,105]}$ & \xmark$_{0/3}$ & $147_{[53,198]}$ & \xmark$_{0/3}$ & $17_{[0,47]}$ & \xmark$_{0/3}$ & $11_{[10,14]}$ & \xmark$_{0/3}$ & $7_{[1,14]}$ & \xmark$_{0/3}$ & $9_{[6,12]}$ \\
26 & \cmark$_{3/3}$ & $77_{[72,81]}$ & \xmark$_{0/3}$ & $63_{[52,74]}$ & \xmark$_{0/3}$ & $4_{[0,7]}$ & \cmark$_{2/3}$ & $48_{[35,64]}$ & \xmark$_{0/3}$ & $9_{[6,13]}$ & \cmark$_{2/3}$ & $32_{[22,41]}$ \\
27 & \cmark$_{2/3}$ & $21_{[20,22]}$ & \xmark$_{0/3}$ & $17_{[15,19]}$ & \cmark$_{1/3}$ & $7_{[0,14]}$ & \xmark$_{0/3}$ & $17_{[15,21]}$ & \xmark$_{0/3}$ & $5_{[4,7]}$ & \xmark$_{0/3}$ & $7_{[6,8]}$ \\
28 & \xmark$_{0/3}$ & $9_{[7,11]}$ & \xmark$_{0/3}$ & $20_{[10,28]}$ & \xmark$_{0/3}$ & $9_{[0,16]}$ & \xmark$_{0/3}$ & $33_{[10,47]}$ & \xmark$_{0/3}$ & $6_{[6,7]}$ & \xmark$_{0/3}$ & $8_{[4,10]}$ \\
29 & \cmark$_{3/3}$ & $48_{[47,50]}$ & \xmark$_{0/3}$ & $24_{[22,28]}$ & \xmark$_{0/3}$ & $21_{[11,33]}$ & \cmark$_{2/3}$ & $38_{[33,47]}$ & \xmark$_{0/3}$ & $8_{[6,10]}$ & \cmark$_{1/3}$ & $26_{[16,42]}$ \\
30 & \xmark$_{0/3}$ & $8_{[5,10]}$ & \xmark$_{0/3}$ & $3_{[1,5]}$ & \xmark$_{0/3}$ & $12_{[5,17]}$ & \xmark$_{0/3}$ & $22_{[11,34]}$ & \xmark$_{0/3}$ & $6_{[5,7]}$ & \xmark$_{0/3}$ & $15_{[9,19]}$ \\
31 & \cmark$_{3/3}$ & $35_{[34,36]}$ & \xmark$_{0/3}$ & $11_{[11,12]}$ & \xmark$_{0/3}$ & $28_{[13,52]}$ & \cmark$_{1/3}$ & $40_{[18,70]}$ & \xmark$_{0/3}$ & $9_{[4,14]}$ & \xmark$_{0/3}$ & $39_{[13,60]}$ \\
32 & \cmark$_{3/3}$ & $109_{[89,144]}$ & \cmark$_{3/3}$ & $69_{[64,73]}$ & \xmark$_{0/3}$ & $21_{[12,35]}$ & \cmark$_{1/3}$ & $38_{[13,69]}$ & \xmark$_{0/3}$ & $10_{[7,13]}$ & \cmark$_{1/3}$ & $27_{[20,34]}$ \\
33 & \cmark$_{2/3}$ & $34_{[22,44]}$ & \xmark$_{0/3}$ & $14_{[13,15]}$ & \xmark$_{0/3}$ & $19_{[18,21]}$ & \xmark$_{0/3}$ & $28_{[16,45]}$ & \xmark$_{0/3}$ & $6_{[5,6]}$ & \xmark$_{0/3}$ & $29_{[11,59]}$ \\
34 & \xmark$_{0/3}$ & $36_{[34,38]}$ & \xmark$_{0/3}$ & $4_{[4,5]}$ & \cmark$_{2/3}$ & $17_{[16,20]}$ & \cmark$_{2/3}$ & $22_{[12,39]}$ & \cmark$_{1/3}$ & $11_{[8,16]}$ & \cmark$_{1/3}$ & $23_{[17,27]}$ \\
35 & \xmark$_{0/3}$ & $32_{[30,34]}$ & \xmark$_{0/3}$ & $6_{[5,7]}$ & \xmark$_{0/3}$ & $25_{[17,39]}$ & \xmark$_{0/3}$ & $16_{[12,24]}$ & \xmark$_{0/3}$ & $7_{[6,8]}$ & \xmark$_{0/3}$ & $15_{[12,17]}$ \\
\midrule
\textbf{Tot.} & \textbf{20} & \textbf{1732} & \textbf{9} & \textbf{2009} & \textbf{5} & \textbf{410} & \textbf{13} & \textbf{1060} & \textbf{7} & \textbf{240} & \textbf{9} & \textbf{598} \\
\bottomrule
\end{tabular}
\end{table*}

\subsection{RQ1: Detection Effectiveness}

Having reduced the analysis space by roughly 31$\times$ on average, we now evaluate whether the remaining functions are sufficient to recover real logic vulnerabilities. To keep the per-repository breakdowns within the column budget, each repository is referred to in the following tables by a short numeric identifier (01--35). The full \texttt{CVE-YYYY-NNNN} label paired with its CWE class is listed in \Autoref{tab:id-mapping} (Appendix~\ref{app:id-mapping}).
\Autoref{tab:comparison-claude} and \Autoref{tab:comparison-gpt} report the end-to-end effectiveness of \tool{} on the 35 repositories in our benchmark, using Opus~4.7 and GPT-5.4 as the reasoning model respectively. We discuss each in turn.
With Opus~4.7, \tool{} detects and explains 20 out of 35 confirmed logic vulnerabilities, producing for each a structured finding that names the security-sensitive sink and the unsatisfied safety condition behind the detection. Detection spans repositories of widely varying size, from a few dozen functions to several hundred, indicating that the approach does not depend on a particular scale of project. Comparative validation accounts for a meaningful share of the reported precision. It removes \textcolor{black}{roughly 25\% of the initial false positives} while leaving every true positive intact, which confirms that validation prunes a substantial fraction of recurring, non-distinctive findings.
With GPT-5.4, \tool{} detects 14 out of 35 vulnerabilities, with validation \textcolor{black}{removing roughly 27 \% of the initial false positives}. The pipeline behavior is consistent with the Opus~4.7 run, high recall over prioritized functions followed by substantial validation pruning, while the set of detected vulnerabilities differs in part. This suggests that the two models surface partially complementary findings under the same framework, and that the gains of \tool{} are not tied to a single reasoning model.

\subsection{RQ2: Comparison with SoTA }

\begin{table}[!t]
\centering
\caption{Comparison of \toolG{} against the GPT function-level baseline. Subscripts indicate stability across three runs. FPs are means with observed ranges. IDs refer to Table~\ref{tab:id-mapping}.}
\label{tab:comparison-gpt}
\renewcommand{\arraystretch}{0.9}
\begin{tabular}{llrlr}
\toprule
& \multicolumn{2}{c}{\toolG{}} & \multicolumn{2}{c}{Func.} \\
\cmidrule(lr){2-3}\cmidrule(lr){4-5}
ID & TPs & FPs & TPs & FPs \\
\midrule
01 & \xmark$_{0/3}$ & $156_{[154,159]}$ & \xmark$_{0/3}$ & $309_{[303,314]}$ \\
02 & \xmark$_{0/3}$ & $54_{[53,55]}$ & \xmark$_{0/3}$ & $46_{[43,51]}$ \\
03 & \xmark$_{0/3}$ & $66_{[63,68]}$ & \xmark$_{0/3}$ & $96_{[87,106]}$ \\
04 & \xmark$_{0/3}$ & $39_{[35,47]}$ & \xmark$_{0/3}$ & $74_{[71,79]}$ \\
05 & \cmark$_{3/3}$ & $529_{[501,550]}$ & \cmark$_{3/3}$ & $814_{[800,836]}$ \\
06 & \cmark$_{3/3}$ & $41_{[39,43]}$ & \xmark$_{0/3}$ & $29_{[25,33]}$ \\
07 & \xmark$_{0/3}$ & $9_{[7,10]}$ & \xmark$_{0/3}$ & $14_{[13,14]}$ \\
08 & \cmark$_{3/3}$ & $14_{[11,16]}$ & \xmark$_{0/3}$ & $20_{[19,22]}$ \\
09 & \cmark$_{3/3}$ & $49_{[48,50]}$ & \xmark$_{0/3}$ & $56_{[54,59]}$ \\
10 & \cmark$_{3/3}$ & $122_{[113,129]}$ & \xmark$_{0/3}$ & $104_{[100,112]}$ \\
11 & \cmark$_{3/3}$ & $104_{[103,106]}$ & \cmark$_{1/3}$ & $84_{[81,87]}$ \\
12 & \xmark$_{0/3}$ & $57_{[56,59]}$ & \xmark$_{0/3}$ & $77_{[70,82]}$ \\
13 & \xmark$_{0/3}$ & $164_{[154,177]}$ & \xmark$_{0/3}$ & $404_{[399,412]}$ \\
14 & \xmark$_{0/3}$ & $123_{[117,130]}$ & \xmark$_{0/3}$ & $261_{[231,280]}$ \\
15 & \xmark$_{0/3}$ & $18_{[17,19]}$ & \xmark$_{0/3}$ & $39_{[35,42]}$ \\
16 & \cmark$_{1/3}$ & $15_{[11,17]}$ & \xmark$_{0/3}$ & $34_{[29,40]}$ \\
17 & \xmark$_{0/3}$ & $75_{[71,82]}$ & \xmark$_{0/3}$ & $121_{[117,127]}$ \\
18 & \xmark$_{0/3}$ & $20_{[18,23]}$ & \xmark$_{0/3}$ & $33_{[31,37]}$ \\
19 & \cmark$_{3/3}$ & $98_{[89,102]}$ & \xmark$_{0/3}$ & $99_{[93,103]}$ \\
20 & \cmark$_{3/3}$ & $53_{[47,60]}$ & \cmark$_{3/3}$ & $44_{[39,49]}$ \\
21 & \xmark$_{0/3}$ & $363_{[355,368]}$ & \xmark$_{0/3}$ & $918_{[886,972]}$ \\
22 & \xmark$_{0/3}$ & $119_{[116,124]}$ & \xmark$_{0/3}$ & $217_{[203,232]}$ \\
23 & \xmark$_{0/3}$ & $49_{[44,55]}$ & \xmark$_{0/3}$ & $60_{[54,65]}$ \\
24 & \cmark$_{3/3}$ & $139_{[132,147]}$ & \cmark$_{3/3}$ & $217_{[209,222]}$ \\
25 & \xmark$_{0/3}$ & $249_{[235,257]}$ & \xmark$_{0/3}$ & $516_{[510,524]}$ \\
26 & \cmark$_{3/3}$ & $81_{[74,91]}$ & \cmark$_{2/3}$ & $132_{[125,140]}$ \\
27 & \xmark$_{0/3}$ & $29_{[26,33]}$ & \xmark$_{0/3}$ & $39_{[36,44]}$ \\
28 & \cmark$_{3/3}$ & $60_{[56,67]}$ & \xmark$_{0/3}$ & $107_{[100,112]}$ \\
29 & \cmark$_{1/3}$ & $123_{[118,128]}$ & \xmark$_{0/3}$ & $110_{[104,114]}$ \\
30 & \xmark$_{0/3}$ & $19_{[13,22]}$ & \xmark$_{0/3}$ & $23_{[22,26]}$ \\
31 & \xmark$_{0/3}$ & $102_{[100,104]}$ & \cmark$_{1/3}$ & $91_{[85,95]}$ \\
32 & \cmark$_{3/3}$ & $130_{[96,150]}$ & \cmark$_{3/3}$ & $152_{[147,154]}$ \\
33 & \xmark$_{0/3}$ & $182_{[172,196]}$ & \cmark$_{3/3}$ & $74_{[71,76]}$ \\
34 & \xmark$_{0/3}$ & $78_{[69,87]}$ & \xmark$_{0/3}$ & $60_{[59,61]}$ \\
35 & \xmark$_{0/3}$ & $77_{[74,80]}$ & \xmark$_{0/3}$ & $42_{[40,44]}$ \\
\midrule
\textbf{Tot.} & \textbf{14} & \textbf{3606} & \textbf{8} & \textbf{5516} \\
\bottomrule
\end{tabular}

\end{table}

We compare \tool{} against two families of baselines, function-level detection (Claude Opus~4.7 and GPT-5.4, each function analyzed without repository context) and agentic detection, in which the model autonomously explores the repository. The agentic baselines are run both over our prioritized function set and over the raw repository without prioritization, using Opus~4.8, Opus~4.7, and Codex (GPT-5.4) in each setting. ~\Autoref{tab:comparison-claude} and \Autoref{tab:comparison-gpt} summarize recall and false positives across all configurations.
\tool{} substantially outperforms every comparison on recall. With Opus~4.7, it recovers \textcolor{black}{20} of 35 vulnerabilities, and with GPT-5.4, 14 of 35, against \textcolor{black}{13} of 35 for the strongest baseline. The gain holds across model families, since the same pipeline lifts both Claude and GPT well above their function-level and agentic counterparts, indicating that the improvement comes from how \tool{} structures and grounds repository context. Recall is also stable across Claude generations. Although ~\Autoref{tab:comparison-claude} reports \tool{} with Opus~4.7, we additionally ran the full pipeline with Opus~4.8 and again recovered 20 of 35 vulnerabilities, indicating the gains are not tied to a particular model version. \textcolor{black}{\tool{}'s detections are also markedly more reproducible across the three repeated runs than any baseline: on average, a confirmed vulnerability is recovered in more than 85\% of the runs for \tool{} with either reasoning model, against at most about 81\% for the function-level baseline and under 60\% for every agentic baseline.}
\tool{} reports more false positives than the agentic baselines, but this reflects the amount of code each method examines. \tool{} and the function-level baseline both analyze every function under consideration, so they surface the full set of candidate findings, \textcolor{black}{whereas the agentic baselines inspect only a subset of functions, applying their own internal prioritization rather than examining the full set, both when given our ranking and when left to explore the raw repository. The ranking appears to nudge the models toward examining more candidates, but they still curtail the search internally, so the extra effort surfaces additional spurious findings without reaching the functions that carry the flaw.}
The right way to weigh these false positives is against the detections they accompany, that is, by discovery yield. \tool{} (Opus~4.7) confirms one vulnerability for roughly every \textcolor{black}{87} reported findings (\textcolor{black}{20 of 1{,}732}, a TP/FPs ratio of \textcolor{black}{0.012}), \textcolor{black}{a yield comparable to one in 82 for the unranked Opus~4.7 agentic baseline (5 of 410, 0.012), which nonetheless recovers only a quarter as many vulnerabilities}. The unranked Opus~4.8 agentic baseline attains a higher ratio (\textcolor{black}{7 of 240, 0.029}) but only by detecting a third as many vulnerabilities, since its narrow coverage keeps both its findings and its detections low. This is the central trade-off: aggressive internal prioritization improves the apparent signal-to-noise ratio while suppressing overall discovery, whereas \tool{} sustains a far higher detection rate at a yield that still keeps triage tractable. Viewed through a bugonomics lens~\cite{pesoli2026demystifying}, \tool{} converts analyst triage effort into confirmed vulnerabilities more effectively than the agentic baselines, improving the return on validation while preserving substantially higher discovery capability. The meaningful comparison is therefore with the function-level baseline, which matches \tool{}'s breadth of analysis. There, the false-positive counts are comparable while \tool{}'s recall is \textcolor{black}{twice as high}, so \tool{} turns the same breadth into many more detections without a corresponding rise in false positives. 
We omit Codex (GPT-5.4) from ~\Autoref{tab:comparison-gpt} because it fails to recover a single vulnerability in either the prioritized or the raw-repository setting, achieving 0 of 35 across both settings, so its column would add nothing to the comparison. 
Across all configurations we tested, no baseline exceeds \textcolor{black}{13} of 35, while \tool{} reaches \textcolor{black}{14-20} of 35, depending on the model. The agentic baselines keep false positives low only by analyzing less code, even when given our ranking, and the function-level baseline matches \tool{}'s breadth but not its recall. \tool{}'s advantage is therefore not a tuning artifact but a property of pairing exhaustive analysis of prioritized code with structured, context-grounded reasoning.

\label{sec:eval:rq3}
\subsection{RQ3: Contribution of Context and FP-Pruning (Ablation)}
\label{sec:eval:rq2}

\begin{table}[!t]
\centering
\caption{Per-CVE detection under context ablations. \emph{$-$Local} removes the local augmentation and \emph{$-$Repo} removes the repository-level context bundle.}
\label{tab:cve-ablation}
\renewcommand{\arraystretch}{0.9}
\begin{tabular}{lcccc}
\toprule
& \multicolumn{2}{c}{\toolO{}} & \multicolumn{2}{c}{\toolG{}} \\
\cmidrule(lr){2-3}\cmidrule(lr){4-5}
\textbf{ID} & $-$Local & $-$Repo & $-$Local & $-$Repo \\
\midrule
01         & \ding{55} & \ding{55} & \ding{55} & \ding{55} \\
02         & \ding{55} & \ding{55} & \ding{55} & \ding{55} \\
03         & \ding{55} & \ding{55} & \ding{55} & \ding{55} \\
04         & \ding{55} & \ding{55} & \ding{55} & \ding{55} \\
05         & \ding{51} & \ding{51} & \ding{51} & \ding{51} \\
06         & \ding{55} & \ding{55} & \ding{55} & \ding{55} \\
07         & \ding{55} & \ding{55} & \ding{55} & \ding{55} \\
08         & \ding{51} & \ding{51} & \ding{51} & \ding{51} \\
09         & \ding{51} & \ding{51} & \ding{51} & \ding{51} \\
10         & \ding{51} & \ding{51} & \ding{51} & \ding{51} \\
11         & \ding{51} & \ding{51} & \ding{51} & \ding{51} \\
12         & \ding{55} & \ding{55} & \ding{55} & \ding{55} \\
13         & \ding{55} & \ding{55} & \ding{55} & \ding{55} \\
14         & \ding{55} & \ding{55} & \ding{55} & \ding{55} \\
15         & \ding{55} & \ding{51} & \ding{55} & \ding{55} \\
16         & \ding{55} & \ding{55} & \ding{55} & \ding{55} \\
17         & \ding{55} & \ding{55} & \ding{55} & \ding{55} \\
18         & \ding{55} & \ding{55} & \ding{55} & \ding{55} \\
19         & \ding{51} & \ding{55} & \ding{51} & \ding{51} \\
20         & \ding{51} & \ding{55} & \ding{51} & \ding{55} \\
21         & \ding{55} & \ding{55} & \ding{55} & \ding{55} \\
22         & \ding{55} & \ding{55} & \ding{55} & \ding{51} \\
23         & \ding{55} & \ding{55} & \ding{55} & \ding{55} \\
24         & \ding{51} & \ding{51} & \ding{55} & \ding{55} \\
25         & \ding{55} & \ding{55} & \ding{55} & \ding{55} \\
26         & \ding{51} & \ding{51} & \ding{51} & \ding{55} \\
27         & \ding{55} & \ding{51} & \ding{55} & \ding{55} \\
28         & \ding{55} & \ding{55} & \ding{55} & \ding{55} \\
29         & \ding{51} & \ding{55} & \ding{55} & \ding{55} \\
30         & \ding{55} & \ding{55} & \ding{55} & \ding{55} \\
31         & \ding{51} & \ding{51} & \ding{55} & \ding{55} \\
32         & \ding{51} & \ding{51} & \ding{51} & \ding{51} \\
33         & \ding{51} & \ding{55} & \ding{55} & \ding{55} \\
34         & \ding{55} & \ding{55} & \ding{55} & \ding{55} \\
35         & \ding{55} & \ding{55} & \ding{55} & \ding{55} \\
\midrule
\textbf{Tot.} & \textbf{13} & \textbf{11} & \textbf{9} & \textbf{8} \\
\bottomrule
\end{tabular}
\end{table}

\begin{table}[!t]
\centering 
\caption{Effect of comparative validation on false positives. \emph{FP}$_{\text{pre}}$ and \emph{FP}$_{\text{post}}$ are FPs before and after auditing.}
\label{tab:ablation-validation}
\setlength{\tabcolsep}{5pt}   
\renewcommand{\arraystretch}{0.9}
\begin{tabular}{l r r r r}
\toprule
& \multicolumn{2}{c}{\toolO{}} & \multicolumn{2}{c}{\toolG{}} \\
\cmidrule(lr){2-3}\cmidrule(lr){4-5}
ID & FP$_{\text{pre}}$ & FP$_{\text{post}}$ & FP$_{\text{pre}}$ & FP$_{\text{post}}$ \\
\midrule
01 & $97_{[97,97]}$ & $80_{[71,86]}$ & $185_{[182,187]}$ & $156_{[154,159]}$ \\
02 & $41_{[38,43]}$ & $35_{[33,37]}$ & $65_{[63,67]}$ & $54_{[53,55]}$ \\
03 & $55_{[52,59]}$ & $49_{[43,53]}$ & $91_{[89,94]}$ & $66_{[63,68]}$ \\
04 & $8_{[7,9]}$ & $6_{[4,7]}$ & $62_{[55,72]}$ & $39_{[35,47]}$ \\
05 & $451_{[444,460]}$ & $342_{[319,356]}$ & $767_{[743,791]}$ & $529_{[501,550]}$ \\
06 & $29_{[28,31]}$ & $22_{[21,23]}$ & $58_{[55,62]}$ & $41_{[39,43]}$ \\
07 & $1_{[1,1]}$ & $1_{[1,1]}$ & $9_{[7,10]}$ & $9_{[7,10]}$ \\
08 & $8_{[7,9]}$ & $6_{[5,7]}$ & $22_{[19,25]}$ & $14_{[11,16]}$ \\
09 & $26_{[22,29]}$ & $23_{[17,28]}$ & $64_{[61,68]}$ & $49_{[48,50]}$ \\
10 & $126_{[120,129]}$ & $103_{[94,108]}$ & $160_{[153,168]}$ & $122_{[113,129]}$ \\
11 & $97_{[91,102]}$ & $79_{[77,83]}$ & $143_{[138,146]}$ & $104_{[103,106]}$ \\
12 & $34_{[32,36]}$ & $25_{[24,28]}$ & $74_{[71,77]}$ & $57_{[56,59]}$ \\
13 & $69_{[64,76]}$ & $49_{[46,51]}$ & $220_{[215,228]}$ & $164_{[154,177]}$ \\
14 & $42_{[35,50]}$ & $29_{[28,30]}$ & $170_{[163,176]}$ & $123_{[117,130]}$ \\
15 & $4_{[3,6]}$ & $3_{[2,4]}$ & $24_{[23,24]}$ & $18_{[17,19]}$ \\
16 & $5_{[5,6]}$ & $5_{[5,6]}$ & $23_{[22,23]}$ & $15_{[11,17]}$ \\
17 & $50_{[46,54]}$ & $41_{[37,44]}$ & $100_{[92,108]}$ & $75_{[71,82]}$ \\
18 & $10_{[8,13]}$ & $9_{[7,13]}$ & $29_{[27,31]}$ & $20_{[18,23]}$ \\
19 & $71_{[71,72]}$ & $54_{[47,58]}$ & $137_{[131,141]}$ & $98_{[89,102]}$ \\
20 & $29_{[21,34]}$ & $20_{[15,24]}$ & $71_{[68,75]}$ & $53_{[47,60]}$ \\
21 & $105_{[93,123]}$ & $79_{[74,82]}$ & $454_{[430,485]}$ & $363_{[355,368]}$ \\
22 & $62_{[58,64]}$ & $52_{[46,57]}$ & $181_{[177,185]}$ & $119_{[116,124]}$ \\
23 & $33_{[30,38]}$ & $25_{[21,29]}$ & $63_{[59,70]}$ & $49_{[44,55]}$ \\
24 & $117_{[114,119]}$ & $92_{[90,95]}$ & $188_{[179,199]}$ & $139_{[132,147]}$ \\
25 & $138_{[125,156]}$ & $94_{[82,105]}$ & $333_{[322,344]}$ & $249_{[235,257]}$ \\
26 & $92_{[89,95]}$ & $77_{[72,81]}$ & $124_{[120,128]}$ & $81_{[74,91]}$ \\
27 & $25_{[25,26]}$ & $21_{[20,22]}$ & $39_{[36,42]}$ & $29_{[26,33]}$ \\
28 & $10_{[8,12]}$ & $9_{[7,11]}$ & $72_{[66,81]}$ & $60_{[56,67]}$ \\
29 & $71_{[66,74]}$ & $48_{[47,50]}$ & $171_{[167,176]}$ & $123_{[118,128]}$ \\
30 & $12_{[7,16]}$ & $8_{[5,10]}$ & $31_{[25,37]}$ & $19_{[13,22]}$ \\
31 & $57_{[55,60]}$ & $35_{[34,36]}$ & $137_{[133,144]}$ & $102_{[100,104]}$ \\
32 & $181_{[141,250]}$ & $109_{[89,144]}$ & $203_{[138,241]}$ & $130_{[96,150]}$ \\
33 & $58_{[48,70]}$ & $34_{[22,44]}$ & $242_{[220,272]}$ & $182_{[172,196]}$ \\
34 & $47_{[44,49]}$ & $36_{[34,38]}$ & $108_{[99,113]}$ & $78_{[69,87]}$ \\
35 & $48_{[46,50]}$ & $32_{[30,34]}$ & $100_{[96,104]}$ & $77_{[74,80]}$ \\
\midrule
\textbf{Tot.} & \textbf{2309} & \textbf{1732} & \textbf{4920} & \textbf{3606} \\
\bottomrule
\end{tabular}
\end{table}
We isolate the contribution of three components of \tool{}, namely the function-level local augmentation (--Local), the repository-level context bundle (--Repo), and the comparative validation that prunes false positives. We ablate each by disabling it and re-measuring detection and false positives.
Removing either context source substantially lowers recall (see \Autoref{tab:cve-ablation}). Under Opus~4.7, detections fall from {\color{black}20} to {\color{black}13} when local augmentation is withheld (--Local) and to {\color{black}11} when the repository bundle is withheld (--Repo). Under GPT-5.4, they fall from {\color{black}14} to {\color{black}9} and {\color{black}8}, respectively. {\color{black}Either ablation reduces the vulnerabilities \tool{} recovers by roughly a third to a half}, confirming that neither source is redundant and that the reasoning stage relies on both the local grounding around a function and the project-wide semantics that situate it. The two sources are also not interchangeable. CVE-2020-10701, for example, is detected only when the repository-level context is present, matching the intuition from our running example, where the safety of the operation could be judged only once the model knew how the surrounding application exercised it. Local augmentation alone leaves that judgment underdetermined.
Removing comparative validation increases false positives at no change to recall, since the stage operates entirely on embeddings and prunes only spurious conditions (see ~\Autoref{tab:ablation-validation}).  With it enabled, false positives fall from {\color{black}2{,}309 to 1{,}732} under Opus~4.7, a {\color{black}25}\% reduction, and from {\color{black}4{,}920 to 3{,}606} under GPT-5.4, a
{\color{black}27}\% reduction in both cases, without dropping any true positives. The effect is larger for GPT-5.4, consistent with its more aggressive flagging behavior, which leaves more spurious conditions for the pruning stage to catch. This stage therefore meaningfully contributes to precision.
Each component contributes a distinct gain. Together, the local and repository context drive the recall improvement that separates \tool{} from the baselines, and comparative validation recovers precision lost to over-flagging. Removing any one degrades a different axis of the result, which is why all three are retained in the full pipeline.

\subsection{RQ4: Cost Analysis}
\label{sec:eval:cost}

We report cost both as token usage and as estimated API spend at list prices. Raw token counts alone are not directly comparable across methods because the agentic baseline relies heavily on prompt caching, where cache reads are billed at a tenth of the input rate, so its raw total is dominated by cheap reuse. \tool{} instead issues uncached prompts whose size is driven by the repository context it embeds and generates very little output. ~\Autoref{tab:cost-summary} reports token usage and ~\Autoref{tab:cost-usd} the resulting cost. Per-CVE breakdowns are in Appendix~\ref{app:cost}.
Three observations summarize the picture. First, \tool{} operates in the same cost regime as the agentic baseline, on the order of a few hundred dollars across all 35 repositories, despite analyzing every prioritized function. The added spend over the function-level baseline is the direct price of the context \tool{} supplies, since both the function-level grounding and the repository-level bundle dominate its prompt size, and it is what the recall gains in RQ2 and RQ3 buy. Second, this spend is well justified by the difference in coverage. \tool{} detects roughly {\color{black}four} times as many vulnerabilities as the agentic baseline, so its cost per detected vulnerability, around \${\color{black}27}, is less than half the agentic baseline's, near \${\color{black}58}. Third, the function-level baseline is cheapest per run but reaches only the most accessible vulnerabilities, detecting {\color{black}9 of 35}, so its low unit cost reflects limited coverage. The comparative validation stage adds no LLM cost, operating entirely on embeddings.

\begin{table}[!t]
\centering
\caption{Aggregate token usage across configurations, in millions of tokens. Cache reports write/read tokens.}
\label{tab:cost-summary}
\renewcommand{\arraystretch}{0.9}
\begin{tabular}{l l r c r r}
\toprule
Model & Config. & Input & Cache (W/R) & Output \\
\midrule
\multirow{3}{*}{Opus 4.7}
& \tool{}      & 101.01 & --          & 0.82  \\
& Func.        & 6.76   & --          & 0.24 \\
& Agentic      & 0.005  & 11.57/320.19 & 1.05  \\
\midrule
\multirow{2}{*}{GPT 5.4}
& \tool{}      & 99.45  & --          & 1.49  \\
& Func.        & 5.97   & --          & 0.75  \\
\bottomrule
\end{tabular}
\end{table}

\begin{table}[!t]
\centering
\caption{Estimated API cost with Opus~4.7 at list prices. Values are approximate.}
\label{tab:cost-usd}
\renewcommand{\arraystretch}{0.9}
\begin{tabular}{l r r r r}
\toprule
Configuration & TPs & Total (\$) & \$/CVE & \$/TPs \\
\midrule
\toolO{}          & \textcolor{black}{20} & $\sim$530 & $\sim$15 & \textcolor{black}{$\sim$27} \\
Opus 4.7 Agentic  &  \textcolor{black}{5} & $\sim$290 & $\sim$8  & \textcolor{black}{$\sim$58} \\
Func.\ Level      &  \textcolor{black}{9} & $\sim$40  & $\sim$1  & \textcolor{black}{$\sim$4}  \\
\bottomrule
\end{tabular}
\end{table}

\FloatBarrier
\section{Discussion}
\label{sec:discussion}

Our aim is not a complete or formally sound vulnerability-analysis framework, but a practical method for recovering repository-specific security invariants under realistic analysis budgets. \tool{} deliberately trades exhaustive coverage for tractability at repository scale, and we make no claim of exhaustive detection, completeness, or formal correctness of the invariants it infers. The results should be read as evidence that repository-grounded invariant recovery can materially improve discovery effectiveness in practice. The rest of this section examines what this trade buys, where its boundaries currently lie, and how the design positions \tool{} to extend past them, with those boundaries understood as trade-offs.

\noindent \textbf{Depth of local grounding.} The local augmentation bundle expands the call graph to depth one, keeping prompts within budget; a safety condition whose justification lives two or more levels away may therefore be judged on incomplete evidence. The repository-level context bundle partially offsets this by supplying project-wide semantics, though as a summary it does not restore arbitrary interprocedural reach. Depth is a tunable parameter, and deeper grounding can be traded against prompt budget where a target repository warrants it.

\noindent \textbf{Reliance on model judgment.} Sink identification and safety-condition derivation come from an LLM, and although the structured output ties each claim to supplied code, the model can still omit a relevant sink, propose a spurious condition, or offer reasoning that is plausible but unfaithful~\cite{turpin2023language}. Comparative validation mitigates this by pruning conditions that recur across peers, but it operates on what the reasoning stage reports and cannot recover a sink the model never surfaced, which is why \tool{} favors structured, evidence-linked output over free-form verdicts: making each claim inspectable is what allows downstream validation and lets a human reviewer audit a finding.

\noindent \textbf{Cost of triage.} \textcolor{black}{%
\tool{} confirms one vulnerability for roughly every 87 reported findings, a ratio that is not negligible even after comparative validation has pruned a substantial share of spurious conditions. We read this as the price of the coverage \tool{} chooses to buy and not as a ceiling on its practical usability: recent advances in LLM-based security automation are substantially reducing the cost of triaging individual findings, with investigations that previously required hours increasingly completed in minutes~\cite{google2026triage, google2026detect}. Recent AI-driven vulnerability discovery also operates at a much larger scale, for context, Project Glasswing reports 23,019 vulnerability candidates, of which 1,596 had been reported to maintainers and 97 patched upstream as of May 22, 2026~\cite{anthropic2026glasswing}, figures not directly comparable to our TP count but indicative of how contained \tool{}'s candidate set is by comparison. Reducing this false-positive burden nonetheless remains an important direction for improving the system's practical usability.
}

\noindent \textbf{Role of prioritization.} Detection examines only the functions that survive prioritization, so this stage bounds what the reasoning stage can ever see. It rests on both keyword heuristics for initial pruning and the agent's ranking of survivors, either of which may overlook code whose relevance is not evident from its name, callees, or structure. We keep the heuristics deliberately high-recall, but surfacing security-relevant code at repository scale without exhaustive analysis remains an open problem in its own right, so we treat prioritization as a pluggable component from which \tool{} stands to benefit directly as the field advances.

\noindent \textbf{Sensitivity to prompt design.} Each LLM stage is driven by prompts, and the space of formulations is effectively unbounded. We iterated on the prompts for prioritization, context extraction, and structured reasoning to obtain reliable, well-structured outputs, but make no claim of optimality: a different phrasing, ordering, or level of instruction could move the results either way, and the numbers we report reflect one carefully chosen, not exhaustively searched, configuration, leaving headroom a more systematic prompt search could likely recover.

\noindent \textbf{Generality of the evaluation.} Our study covers 35 real-world C and C++ repositories with confirmed CWE-200 and CWE-284 vulnerabilities. Labeled logic vulnerabilities of this kind are scarce, so assembling even this many is non-trivial and substantial relative to what is available for the class, though the sample remains small in absolute terms, limiting statistical power and possibly not capturing the full diversity of C and C++ conventions. Results may likewise not transfer directly to other languages, proprietary codebases with different norms, or vulnerability classes beyond the two we target. We see broadening across languages and CWE classes as the natural next step, and the language-agnostic structure of the pipeline, prioritize, augment, reason, validate, is intended to make that extension straightforward.
\section{Related Work}
\label{sec:related}

The earliest systematic treatments of logic flaws come from the web domain, deriving invariants from observed executions or behavioral models learned from request traces, and specializing this idea to application workflows and e-commerce checkout logic~\cite{felmetsger2010toward, pellegrino2014toward, deepa2018detlogic, li2013logicscope, sun2014detecting}. These works establish that the violated property is defined by the application, but the signals they rely on, such as sessions, forms, and navigation structure, have no analog in systems software.
A separate tradition infers correctness rules from the code itself and treats violations as defects. Engler et al. extract implicit programming rules from systems code and surface statistical outliers as candidate bugs~\cite{10.1145/502034.502041}, and Chucky identifies missing security checks by contrasting a function with structurally comparable code and reporting the omissions that stand out~\cite{10.1145/2508859.2516665}. For the invariants we target, none of these anchors is available, and project-specific rules can encode them only at a manual cost that does not scale across repositories. \tool{} retains the anomaly-based intuition of the missing-check line of work, but applies it over structured, model-produced findings.

A growing body of work applies language models to vulnerability detection. Many systems operate at the function level, presenting a single function and asking for a label or report, optionally augmented with external knowledge such as distilled vulnerability descriptions or CWE references retrieved at inference time to steer the judgment~\cite{10.1109/ICSE55347.2025.00038, chen2025reasoning, du2024vul, 10620804}. A complementary direction couples models with program analysis so that reasoning is anchored to extracted facts, whether taint specifications driving a static analyzer over the whole repository, model queries embedded inside a static-analysis loop, or a code property graph compressed into compact slices that let a model reason over segments spanning multiple functions~\cite{li2025iris, llift, chapman2024interleaving, lekssays2025llmxcpg}. These systems anchor on a tainted source, a specification, or a graph path, the kind of scaffold that logic flaws do not provide.
A further line moves detection from the function to the repository. Benchmarking efforts show that inter-procedural detection is a distinct and largely open problem, and supply call-graph dependencies as the bridge from intra- to inter-procedural reasoning~\cite{wen2026function}. Autonomous auditing agents instead traverse a repository to locate and explain defects~\cite{guo2025repoaudit}. Both rely on a dependency or anchor to follow, which the CWE-284 and CWE-200 issues lacks, leaving open-ended search prone to both wasted budget and drift. \tool{} instead fixes a bounded sequence of steps: it narrows the candidate set, attaches targeted evidence, and prunes by comparison. The open question is therefore not whether frontier models can surface real vulnerabilities given sufficient scaffolding and compute~\cite{carlini2026mythos}, but how to harness that capability at repository scale without making the analysis prohibitively expensive~\cite{pesoli2026demystifying}. To the best of our knowledge, \tool{} is the first system to propose a cost-effective repository-level solution to this challenge for logic vulnerabilities.
\section{Conclusion}
\label{sec:conclusion}

We presented \tool{}, a repository-level framework for detecting logic vulnerabilities by grounding LLM reasoning in codebase-wide context. \tool{} follows a five-stage pipeline that combines function prioritization, context-grounding, structured reasoning, comparative validation, and reporting to make LLM analysis both evidence-grounded and cost-aware. It prioritizes security-relevant functions, grounds each in local and repository-level evidence, derives sinks and safety conditions, validates candidate findings against similar sinks, and reports retained violations with supporting evidence. Our evaluation on 35 real-world C and C++ repositories with confirmed CWE-284 and CWE-200 vulnerabilities shows that \tool{} detects and explains 20 vulnerabilities, substantially outperforming function-level LLM analysis and frontier-agentic baselines while maintaining comparable token usage and cost budgets. These results suggest that repository-specific grounding and staged validation are key to efficiently applying frontier LLM capabilities to logic vulnerability detection.

\bibliographystyle{IEEEtran}
\bibliography{references}

\appendix
\section{Ethical Considerations}
This work focuses on the automated detection of logical vulnerabilities in software repositories. Our benchmark consists exclusively of previously disclosed vulnerabilities with public CVE identifiers. While automated vulnerability-detection tools pose inherent dual-use risks, our tool is intended to support defensive security analysis, targets context-dependent logical flaws, and does not generate weaponized exploits. This research does not involve human subjects or user data.

\section{Open Science}
In the spirit of open science, we provide the artifacts underlying our experiments at the following anonymous link:
\begin{itemize}
\item \url{https://anonymous.4open.science/r/AntaeusUsenix-8B6F}
\end{itemize}
The repository includes all scripts needed to run \tool{}, the full set of prompts, and documentation covering both how to run \tool{} and the benchmark repositories used in our evaluation. We are currently polishing the full implementation of \tool{} and plan to release it as open source.

\section{Structured Report Schema}
\label{app:report-schema}

\Autoref{lst:report-schema} shows the complete schema of the structured report produced by \tool{}. Each entry identifies an analyzed function and the security-sensitive sinks detected within it. For every sink, the report records the conditions required for the operation to be secure, whether each condition is locally satisfied, and the corresponding code-grounded justification. A function is included in the report only if at least one required condition is not satisfied locally.

\begin{codelisting}[t]
\begin{lstlisting}[style=cstyle,basicstyle=\ttfamily\footnotesize]
{
  "cwe": <weakness class>,
  "function_name": <name>,
  "file": <path>,
  "lines": <range>,
  "function id": <function id>
  "sinks": [
    {
      "sink_id": <security-sensitive operation>,
      "sink_description": <why it crosses a trust or exposure boundary>,
      "required_conditions": [
        {
          "id": <invariant name>,
          "description": <invariant the operation depends on>,
          "locally_satisfied": <true | false>,
          "justification": <code evidence supporting the judgment>
        }
      ]
    }
  ]
}
\end{lstlisting}
\caption{Schema of a structured report entry. A function appears only when at
least one sink has a required condition with \texttt{locally\_satisfied} set to
false; that condition and its justification constitute the reported
vulnerability evidence.}
\label{lst:report-schema}
\end{codelisting}

\section{Repository-Identifier Mapping}
\label{app:id-mapping}

~\Autoref{tab:id-mapping} shows the mapping between the two-digit repository identifiers used throughout the evaluation in Section~\ref{sec:eval} and their corresponding CVEs. IDs~01--28 refer to the ReposVul CVEs, ordered by CVE number, while IDs~29--35 correspond to the seven post-cutoff CVEs. The table also reports the CWE class associated with each repository.

\begin{table}[t!]
\centering
\caption{Mapping between the repository identifiers used in the evaluation and their corresponding CVEs and CWE classes.}
\label{tab:id-mapping}

\begin{tabular}{l l l}
\toprule
\textbf{ID} & \textbf{CVE} & \textbf{CWE} \\
\midrule
01 & CVE-2011-3177  & CWE-200 \\
02 & CVE-2014-4668  & CWE-284 \\
03 & CVE-2014-9773  & CWE-284 \\
04 & CVE-2015-2141  & CWE-200 \\
05 & CVE-2016-10030 & CWE-284 \\
06 & CVE-2016-10130 & CWE-284 \\
07 & CVE-2016-3698  & CWE-284 \\
08 & CVE-2016-5104  & CWE-284 \\
09 & CVE-2016-6255  & CWE-284 \\
10 & CVE-2017-5940  & CWE-284 \\
11 & CVE-2017-5985  & CWE-284 \\
12 & CVE-2017-6594  & CWE-284 \\
13 & CVE-2018-19045 & CWE-200 \\
14 & CVE-2018-20145 & CWE-200 \\
15 & CVE-2019-12209 & CWE-200 \\
16 & CVE-2019-12210 & CWE-200 \\
17 & CVE-2019-12589 & CWE-284 \\
18 & CVE-2019-25016 & CWE-200 \\
19 & CVE-2019-3816  & CWE-200 \\
20 & CVE-2020-10701 & CWE-284 \\
21 & CVE-2020-14093 & CWE-200 \\
22 & CVE-2020-14976 & CWE-200 \\
23 & CVE-2020-15078 & CWE-284 \\
24 & CVE-2020-29074 & CWE-284 \\
25 & CVE-2020-35517 & CWE-284 \\
26 & CVE-2022-24976 & CWE-284 \\
27 & CVE-2022-25643 & CWE-200 \\
28 & CVE-2023-27478 & CWE-200 \\
29 & CVE-2026-21636 & CWE-284 \\
30 & CVE-2026-32814 & CWE-200 \\
31 & CVE-2026-34766 & CWE-284 \\
32 & CVE-2026-45313 & CWE-284 \\
33 & CVE-2026-6429  & CWE-200 \\
34 & CVE-2026-67970 & CWE-284 \\
35 & CVE-2026-67972 & CWE-200 \\
\bottomrule
\end{tabular}

\end{table}

\section{Per-Repository Prioritization Statistics}
\label{app:pruning}

\Autoref{tab:pruning-stats-full} gives the per-repository breakdown behind the aggregate figures of \Autoref{tab:pruning-summary}. For each repository it reports, separately for the CWE-284 and CWE-200 passes, the files retained after heuristic pruning, the in-scope functions in those files, and the functions saved by the agent after ranking. The shared \emph{Files} column is the total source and header file count, identical across both passes since both operate on the same repository.

\begin{table}[t!]
\centering
\caption{Per-CVE pruning stats for CWE-284 and CWE-200 passes. \emph{Files} is the total source files, \emph{Ret.} denotes files retained after heuristic pruning; \emph{Scope} and \emph{Anal.} denote in-scope and analyzed functions. IDs refer to Table~\ref{tab:id-mapping}.}
\label{tab:pruning-stats-full}
\setlength{\tabcolsep}{3pt}
\vspace{1mm}
\begin{tabular}{l r rrr rrr}
\toprule
& & \multicolumn{3}{c}{CWE-284} & \multicolumn{3}{c}{CWE-200} \\
\cmidrule(lr){3-5}\cmidrule(lr){6-8}
ID & Files & Ret. & Scope & Anal. & Ret. & Scope & Anal. \\
\midrule
01     &    2962 &   220 &    612 &    41 &   232 &   1180 &   139 \\
02     &    1760 &   127 &    612 &    34 &   327 &    760 &    15 \\
03     &    1830 &   156 &   1180 &    78 &   380 &   1180 &    15 \\
04     &     407 &    33 &    295 &    26 &   103 &    910 &   107 \\
05     &    2668 &   937 &   8195 &   661 &  1041 &  10045 &   120 \\
06     &    5317 &   255 &   1884 &    42 &   530 &   2105 &    21 \\
07     &      48 &     2 &     71 &    10 &     5 &     75 &     5 \\
08     &      50 &     5 &     52 &    12 &     8 &     48 &     6 \\
09     &     551 &    52 &    285 &    61 &   113 &    466 &    12 \\
10     &     408 &    71 &    360 &   125 &    75 &    217 &    12 \\
11     &     469 &   112 &    905 &    88 &   154 &    620 &    16 \\
12     &    2473 &   335 &   3200 &    50 &  1131 &   7200 &    22 \\
13     &     329 &   104 &    720 &    47 &   139 &   1225 &   128 \\
14     &     728 &    80 &    399 &    43 &   118 &    450 &   174 \\
15     &      58 &     5 &     25 &     8 &     7 &     28 &     8 \\
16     &      58 &     5 &     24 &     5 &     7 &     27 &    12 \\
17     &    1416 &   135 &    470 &    65 &   146 &    420 &    42 \\
18     &      62 &     8 &     28 &    20 &    19 &     38 &    11 \\
19     &     542 &    56 &    420 &    73 &   180 &    545 &    34 \\
20     &    9890 &   477 &  12882 &   250 &   692 &   5400 &    15 \\
21     &     441 &   119 &    990 &   133 &   220 &   2054 &  1079 \\
22     &      77 &    25 &    180 &    39 &    30 &    180 &    83 \\
23     &     499 &   138 &   1100 &    38 &   247 &   1011 &    16 \\
24     &     223 &   100 &    760 &   102 &   101 &    470 &    34 \\
25     &    7915 &  1557 &  29128 &   124 &  3500 &  44032 &   267 \\
26     &    2297 &   331 &   1450 &   106 &   440 &   1180 &    18 \\
27     &      76 &    21 &    137 &    30 &    24 &    120 &    11 \\
28     &     588 &    82 &    588 &     9 &   192 &    820 &   117 \\
29     &     472 &   126 & 3{,}457 &    88 &   334 & 4{,}800 &   130 \\
30     &     728 &    67 & 2{,}919 &    53 &   333 & 1{,}500 &    40 \\
31     & 2{,}818 &   186 & 2{,}725 &    31 &   403 & 2{,}725 &   193 \\
32     & 1{,}645 &   339 & 5{,}415 &   183 &   476 & 5{,}531 &    36 \\
33     & 4{,}311 &   336 & 3{,}252 &   165 &   672 & 3{,}200 &   124 \\
34     &     182 &    21 &     89 &    56 &    62 &    405 &    60 \\
35     &     164 &    45 &    523 &    81 &   106 &    560 &    13 \\
\midrule
\textbf{Tot.} & \textbf{54{,}462} & \textbf{6{,}668} & \textbf{85{,}332} & \textbf{2{,}977}
               & \textbf{12{,}547} & \textbf{101{,}527} & \textbf{3{,}135} \\
\bottomrule
\end{tabular}
\end{table}

\section{Per-CVE Token Count}
\label{app:cost}
This appendix gives the per-repository token breakdown behind the aggregate figures reported in \Autoref{sec:eval:cost}. \Autoref{tab:cost-percve} reports the prompt (\emph{In}) and generated (\emph{Out}) tokens for \tool{} and the function-level baseline, under both \toolO{} and \toolG{}, summed over the CWE-284 and CWE-200 passes. \Autoref{tab:cost-opus47-agentic} reports the corresponding usage for the Opus~4.7 agentic baseline, broken down by token category because it relies on prompt caching.
\begin{table}[t!]
\centering
\caption{Per-CVE token count for \tool{}, summed over the CWE-284 and CWE-200 passes. All values in thousands of tokens. \emph{In}/\emph{Out} are prompt and generated tokens; \emph{Func.} is the function-level baseline.}
\label{tab:cost-percve}
\setlength{\tabcolsep}{3pt}
\vspace{1mm}
\begin{tabular}{l rr rr rr rr}
\toprule

& \multicolumn{2}{c}{\toolO{}} & \multicolumn{2}{c}{Func.\textsuperscript{Opus4.7}} & \multicolumn{2}{c}{\toolG{}} & \multicolumn{2}{c}{Func.\textsuperscript{GPT5.4}} \\
\cmidrule(lr){2-3}\cmidrule(lr){4-5}\cmidrule(lr){6-7}\cmidrule(lr){8-9}
ID & In & Out & In & Out & In & Out & In & Out \\
\midrule
01     &  2{,}077 &   21 &      166 &   17 &  1{,}972 &   45 &      126 &   36 \\
02     &  1{,}045 &   14 &       60 &    5 &  1{,}034 &   18 &       55 &    6 \\
03     &  1{,}661 &   25 &      104 &    7 &  1{,}651 &   34 &       99 &   15 \\
04     &  2{,}083 &    2 &      115 &    1 &  1{,}994 &   18 &       81 &   10 \\
05     & 13{,}632 &  184 &      962 &   43 & 13{,}550 &  281 &      920 &  114 \\
06     &  1{,}021 &   12 &       59 &    1 &  1{,}006 &   19 &       52 &    3 \\
07     &      208 &    1 &       13 &    0 &      204 &    3 &       11 &    2 \\
08     &      294 &    2 &       16 &    2 &      290 &    6 &       14 &    3 \\
09     &  1{,}255 &   12 &       93 &    2 &  1{,}247 &   25 &       89 &    9 \\
10     &  2{,}344 &   42 &      128 &    7 &  2{,}335 &   48 &      124 &   16 \\
11     &  1{,}796 &   34 &      102 &    7 &  1{,}785 &   41 &       97 &   13 \\
12     &  1{,}409 &   20 &      128 &    5 &  1{,}393 &   30 &      121 &   11 \\
13     &  3{,}495 &   19 &      222 &    7 &  3{,}407 &   59 &      177 &   44 \\
14     &  4{,}498 &   18 &      301 &    6 &  4{,}376 &   50 &      239 &   30 \\
15     &      286 &    3 &       26 &    1 &      281 &    7 &       24 &    5 \\
16     &      326 &    3 &       26 &    1 &      318 &    7 &       22 &    5 \\
17     &  1{,}895 &   27 &      127 &    6 &  1{,}865 &   41 &      112 &   15 \\
18     &      562 &    6 &       28 &    0 &      554 &   11 &       24 &    3 \\
19     &  1{,}902 &   21 &       99 &   11 &  1{,}879 &   38 &       88 &   14 \\
20     &  4{,}159 &   13 &      235 &    1 &  4{,}156 &   22 &      234 &    7 \\
21     & 20{,}408 &   39 &  1{,}232 &   10 & 19{,}797 &  157 &      922 &  108 \\
22     &  2{,}096 &   15 &      116 &    5 &  2{,}038 &   47 &       87 &   26 \\
23     &  1{,}184 &   15 &      107 &    5 &  1{,}173 &   21 &      102 &    7 \\
24     &  3{,}060 &   42 &      273 &   26 &  3{,}036 &   56 &      260 &   28 \\
25     &  6{,}843 &   40 &      552 &    9 &  6{,}657 &   95 &      458 &   65 \\
26     &  2{,}141 &   38 &      149 &   12 &  2{,}128 &   49 &      143 &   21 \\
27     &      741 &    8 &       33 &    3 &      733 &   12 &       29 &    5 \\
28     &  1{,}548 &    2 &      104 &    2 &  1{,}478 &   23 &       78 &   13 \\
29     &  2{,}608 &   24 &      157 &    5 &  2{,}623 &   37 &      158 &   20 \\
30     &  1{,}290 &    4 &      105 &    1 &  1{,}295 &    6 &      106 &    4 \\
31     &  3{,}057 &   17 &      206 &    4 &  3{,}082 &   28 &      208 &   18 \\
32     &  3{,}283 &   48 &      247 &   20 &  3{,}273 &   58 &      249 &   36 \\
33     &  4{,}160 &   25 &      307 &    4 &  4{,}183 &   57 &      309 &   16 \\
34     &  1{,}542 &   10 &       88 &    2 &  1{,}550 &   22 &       89 &   12 \\
35     &  1{,}104 &   11 &       65 &    2 &  1{,}109 &   22 &       65 &    7 \\
\midrule
\textbf{Tot.} & \textbf{101{,}013} & \textbf{817} & \textbf{6{,}758} & \textbf{240}
               & \textbf{99{,}453} & \textbf{1{,}493} & \textbf{5{,}969} & \textbf{749} \\
\bottomrule
\end{tabular}
\end{table}

\begin{table}[t!]
\centering
\caption{Token usage per CVE for Opus 4.7 Agentic, summed over the CWE-200 and CWE-284 passes. Columns marked with K are in thousands of tokens. Total is the row-wise sum of input, cache-write, cache-read, and output tokens.}
\label{tab:cost-opus47-agentic}
\setlength{\tabcolsep}{4pt}
\vspace{1mm}
\begin{tabular}{l r r r r r}
\toprule
ID & Input & CacheW$_K$ & CacheR$_K$ & Out$_K$ & Total$_K$ \\
\midrule
01     &   200 &    391 &   11{,}833 &    35 &   12{,}259 \\
02     &   181 &    404 &   11{,}363 &    18 &   11{,}785 \\
03     &   228 &    798 &   20{,}961 &    30 &   21{,}790 \\
04     &   121 &    274 &    4{,}979 &    25 &    5{,}277 \\
05     &   194 &    375 &   11{,}838 &    19 &   12{,}232 \\
06     &   208 &    351 &   11{,}064 &    37 &   11{,}452 \\
07     &    53 &    167 &        769 &     5 &        941 \\
08     &    52 &    126 &        693 &     8 &        826 \\
09     &   148 &    316 &    8{,}096 &    15 &    8{,}427 \\
10     &   148 &    335 &    9{,}842 &    27 &   10{,}204 \\
11     &   155 &    269 &    6{,}839 &    17 &    7{,}125 \\
12     &   222 &    439 &   14{,}773 &    43 &   15{,}255 \\
13     &   159 &    236 &    6{,}481 &    27 &    6{,}744 \\
14     &   162 &    295 &    7{,}626 &    28 &    7{,}950 \\
15     &    51 &     97 &        557 &    12 &        666 \\
16     &    53 &     99 &        626 &     6 &        730 \\
17     &   242 &    346 &   14{,}957 &    37 &   15{,}339 \\
18     &    62 &    130 &    1{,}007 &    19 &    1{,}156 \\
19     &   150 &    352 &    9{,}722 &    16 &   10{,}091 \\
20     &   278 &    422 &   18{,}380 &    42 &   18{,}845 \\
21     &   236 &    391 &   15{,}290 &    25 &   15{,}706 \\
22     &    71 &    222 &    2{,}272 &    17 &    2{,}512 \\
23     &   194 &    329 &   11{,}306 &    31 &   11{,}666 \\
24     &   187 &    261 &    8{,}268 &    32 &    8{,}562 \\
25     &   247 &    320 &   14{,}722 &    47 &   15{,}090 \\
26     &   245 &    540 &   19{,}277 &    37 &   19{,}855 \\
27     &    75 &    197 &    1{,}885 &    10 &    2{,}092 \\
28     &   124 &    332 &    6{,}321 &    21 &    6{,}674 \\
29     &   171 &    520 &   11{,}073 &    56 &   11{,}649 \\
30     &   106 &    211 &    4{,}981 &    33 &    5{,}225 \\
31     &   197 &    388 &   13{,}436 &    57 &   13{,}881 \\
32     &   204 &    723 &   19{,}514 &    86 &   20{,}323 \\
33     &   199 &    310 &   11{,}704 &    65 &   12{,}079 \\
34     &    90 &    327 &    3{,}220 &    31 &    3{,}578 \\
35     &   112 &    284 &    4{,}524 &    38 &    4{,}846 \\
\midrule
\textbf{Tot.} & \textbf{5{,}525} & \textbf{11{,}576} & \textbf{320{,}197} & \textbf{1{,}053} & \textbf{332{,}832} \\
\bottomrule
\end{tabular}
\end{table}

\end{document}